\documentclass[12pt]{iopart}

\expandafter\let\csname equation*\endcsname\relax

\expandafter\let\csname endequation*\endcsname\relax
\usepackage{iopams}  
\usepackage{amsmath}
\usepackage[colorlinks=true,linkcolor=black, citecolor=blue, urlcolor=blue]{hyperref}
\usepackage{cite}
\usepackage{graphicx}
\usepackage{multirow}
\usepackage{subfig}
\usepackage{mathrsfs}
\usepackage{booktabs}
\begin{document}

\title[Strong Lensing and Quasinormal modes of a Black Hole with a Global Monopole]{Strong Lensing and Quasinormal modes of a Black Hole with a Global Monopole}
%
\author{Irengbam Roshila Devi$^{1a}$,Ningthoujam Media$^{1b}$ Yenshembam Priyobarta Singh$^{1c}$ \& Telem Ibungochouba Singh$^{1d}$}
\address{$^1$Department of Mathematics, Manipur University, Canchipur, Imphal, 795003, Manipur, India }
\ead{$^a$roshilairengbam@gmail.com \& $^b$medyaningthoujam@gmail.com \& $^c$priyoyensh@gmail.com \& $^d$ibungochouba@rediffmail.com (corresponding author) }
\vspace{10pt}
\begin{abstract}
In this paper, we investigate various key aspects of a static and spherically symmetric black hole with a global monopole. Firstly, we analyze the deflection angle in the strong field limit  of a spacetime with a global monopole. It shows that the angle of deflection increases when the two characteristic parameters for the monopole configuration, $\eta$ and $\lambda$, increase. The influence of the global monopole parameter on the lensing observables and the black hole shadow is studied. This shows that a larger monopole parameter corresponds to larger shadow radii.  The dynamics of timelike geodesics is also investigated in the spacetime. General circular orbits and the innermost stable circular orbits (ISCO) of timelike particles are discussed, highlighting that the monopole parameter significantly affects the circular orbits and the ISCO. In particular, it is observed that the radius of ISCO rises monotonically with $\eta$. In addition, the Lyapunov exponent is used to analyze the stability of timelike geodesics.  We also determine the quasinormal modes under electromagnetic perturbations for different values of $\eta$. Our findings indicate that increasing the monopole parameter gives rise to electromagnetic perturbations with more slowly damped oscillations. To further validate the derived quasinormal mode spectrum, we discuss the  time-domain evolution of the perturbation, confirming the presence of the characteristic quasinormal ringing followed by late-time power-law tails.   Finally, comparison with the EHT observations of M87* and Sgr A* determines the parameter combinations of $\eta$ and $\lambda$ compatible with current observations, and the observational prospects of the strong-lensing observables are discussed.

\end{abstract}

%
\vspace{2pc}

\section{Introduction}
\label{intro}
\par

One of the key theoretical outcomes of general relativity is the phenomenon of gravitational lensing in which the curvature of spacetime caused by a massive object (such as a galaxy, cluster of galaxies, or black hole) deflects the path of light \cite{Einstein1936, Erdem 2026(8) }. The phenomenon occurs when a massive object interposes itself between a far-off light source (like another galaxy or a star) and an observer. The effect was initially detected in sunlight deflection and subsequently in quasar lensing by galaxies. Evolving from the basic prediction of light deflection, gravitational lensing has become a cornerstone technique in both astrophysics and cosmology \cite{Adler, Ovgun2025}. The utility of gravitational lensing extends to mass measurements of galaxies and clusters, determination of the Hubble constant, and the study of dark objects such as black holes or massive compact halo objects and dark energy \cite{Virbhadra2024, Ovgun2025}. The theory of gravitational lensing was first formulated using the weak field approximation, an approach that has proven highly effective in accounting for astronomical observations \cite{Schneider}. The weak approximation is invalid near compact stars, where light can orbit them multiple times before escaping. In a recent study, Virbhadra and Ellis \cite{Virbhadra2000} investigated gravitational lensing within the strong field regime and derived the corresponding lens equation.  Refs. \cite{Perlick2004, Schmidt2008} showed that gravitational lensing research significantly enhanced our comprehension of spacetime. Darwin \cite{Darwin1959} was one of the first to use the principles of lensing to study Schwarzschild black holes. Luminet \cite{Luminet1979} extended the lensing near the photon sphere by deriving the logarithmic approximation for light, a formulation now called the strong deflection limit. Virbhadra and Ellis \cite{Virbhadra2000, Virbhadra2002} conducted foundational theoretical research on the rings and magnified relativistic images produced by Schwarzschild black holes. Ref. \cite{Frittelli2000} later investigated exact, integral-form solutions to the lens equation. Based on earlier work, Bozza \cite{Bozza2001, Bozza2002, Bozza2003, Bozza2007} and Tsukamoto \cite{Tsukamoto2016, Tsukamoto2017} developed analytical strong-lensing techniques for static spherically symmetric spacetimes, successfully calculating the positions of higher-order images and magnifications for Schwarzschild, rotating black holes and generic spherically symmetric. Torres \cite{Eiroa2002} analytically calculated the positions and magnifications of relativistic images for Reissner-Nordström black holes. The strong gravitational field continues to be an important focus of research. Recent studies have examined lensing effects from different black holes \cite{Chen2009, Sarkar2006, Javed2019, Shaikh2019} as well as modifications to the Schwarzschild geometry \cite{Shaikh2019b, Babar2021}, including those in higher curvature gravity \cite{Narzi2021, Kumar2020}. In particular, strong lensing and other compact objects produce observable signatures such as shadows, photon rings and relativistic images. This field has been extended in the context of string theories \cite{Bhadra2003, Molla2024},  Lorentz violating frameworks \cite{Filho 2026, Filho 2026a, Pereira 2026a, Filho 2025} and topological charge black holes \cite{Pereira 2026, Soares 2026}.

One of the theoretical consequences of General Relativity (GR) is the bending of light by gravity, which leads to the formation of a black hole's shadow. For this reason, a detailed examination of null geodesics under those spacetimes is crucial. The region close to a black hole allows for circular photon paths, called light rings. These rings define the boundary of a dark area in the sky, which is observed as the black hole shadow. Falcke et al. \cite{Falcke2000} were the first to suggest the idea of observing the black hole shadow indicating that mass and spin are the black hole's key parameters. Although mass classification is resolved, measuring spin is still an ongoing issue. Black hole shadow analysis is considered a promising tool for estimating the spin of rotating black holes \cite{Takahashi}. Recent observations of gravitational waves from merging black holes \cite{Abbott} and images of black hole shadows in the Milky Way \cite{Akiyama2022} and M87* \cite{Akiyama2019} have increased interest in studying black hole spacetimes.  The study of null geodesics and their related optical phenomena, particularly black hole shadows, continues to be a vibrant field of research within rotating black hole spacetimes \cite{He, Belhaj}. The shadow of a Schwarzschild black hole was first analyzed by Synge \cite{Synge1966}. Later, Luminet developed a method for calculating the angular radius of such a shadow \cite{Luminet1979}. The shape of a black hole's shadow depends on its spin: it is circular for a non-rotating black hole, but becomes elongated along the axis of rotation when the black hole is spinning, due to frame-dragging. Based on the distinctive spots of the shadow boundary of the Kerr black hole, Hioki and Maeda \cite{Maeda2009} introduced two observables. The angular size of the black hole shadow is described by one of these observables, while its deformation or departure from a complete circle is described by the other. Refs. \cite{Bambi2009, Tsukamoto2014, Afrin2021, Afrin2022, Okyay2022} have analysed the angular size and form of shadows for various rotating black holes.  Additional related studies can be found in Refs. \cite{Tsukamoto2018, Ovgun2020,Banerjee2020, Konoplya2021, Guo2021}.

However, examining the formation and behavior of topological defects (like cosmic strings and monopoles) has recently become a highly active area in modern physics because their unique properties lead to many unusual physical phenomena.
 According to Grand Unified Theories, global monopoles represent a distinct category of topological defects that are believed to have been generated during the early universe through  the spontaneous breaking of global $O(3)$ gauge symmetry down to $U(1)$ \cite{Kibble, Vilenkin}. An extensive study of their gravitational properties showed that it introduces a solid deficit angle \cite{Erdem 2026(2)}. This fundamentally alters the topology of a black hole, resulting in significant physical differences between monopole and non-monopole black holes \cite{Pan2008, Chen2010, Sorou2020}. 

It is widely recognized that a perturbed black hole exhibits damped oscillations, described by specific complex eigen values of the wave equations, known as quasinormal mode (QNM) frequencies.  They characterize the ringdown phase of gravitational-wave signals and encode the mass, spin and environmental influences of the black hole. QNM frequencies are determined by the effective potential \cite{Erdem 2026(3)} and spacetime structure, making them valuable for testing strong-field gravity and identifying deviations from standard solutions. These modes arise from solutions to perturbed equations with specific boundary conditions: purely outgoing waves exist at infinity and purely ingoing waves exist at the black hole horizon \cite{Chandrasekhar1975a}. The emission of gravitational radiation prevents black holes from having normal oscillation modes. Hence, their characteristic frequencies are quasinormal, existing as complex numbers \cite{Kokkotas}. For QNMs, the oscillation frequency is given by the real part of the frequency, whereas the damping rate is given by the imaginary part \cite{Moss2002}.  Many researchers have developed various methods to find QNMs for different types of black holes such as the Wentzel–Kramers–Brillouin (WKB) approximation \cite{Schutz1985,Iyer1987a,Iyer1987b,Roshila,Jayasri2025}, the Frobenius method \cite{Konoplya2011}, the Poschl–Teller fitting method \cite{Ferrari,PriyoEPJC, Ningthoujam2025},  the Continued fraction method (Leaver’s method)\cite{Leaver}, the Mashhoon method \cite{Mashhoon}, the Asymptotic Iteration method (AIM) \cite{Cho2010, Pong2019,Gogoi,Yenshembam b2025} .

The geodesic structure of spacetime reveals how particles and light move through curved space. In black hole spacetimes, light paths determine observable features like  lensing effects, shadows, and photon spheres, while time-like geodesics govern orbital motion and precession effects. The time-like geodesic structure of Schwarzschild spacetime is thoroughly examined by Chandrasekhar \cite{Chandrasekhar2010}, which includes graphical representations of both bound and unbound orbits. This analysis is later extended to the Schwarzschild anti-de Sitter spacetime in \cite{Cruz}. They examined radial as well as non-radial paths for both time-like and null geodesics. Furthermore, they showed that this black hole's geodesic configuration allows for new types of motion not found in Schwarzschild spacetime. The geodesic structure of Schwarzschild spacetime is also discussed in \cite{Berti2014}. Ref. \cite{Gibbons2016} introduced the Jacobi metric for paths followed by massive particles moving freely under gravity in unchanging, non-rotating spacetimes, showing that massive particle motion follows an energy-dependent Riemannian metric on spatial slices. When the mass approaches zero, this metric becomes equivalent to the optical (Fermat) metric, which does not depend on energy with specific application to Schwarzschild black holes. Reference \cite{Chanda2017} extended this framework to stationary metrics and, via the Eisenhart-Duval lift \cite{Eisenhart}, formulates a Jacobi-Maupertuis metric for time-dependent cases.

 Another key feature in the study of black holes is geodesic stability which governs the behaviour of photons and test particles in the vicinity of black holes \cite{Erdem 2026(9)}. One prominent method for evaluating geodesic stability involves computing the Lyapunov exponent, which characterizes the average rate at which two initially close trajectories in phase space diverge or converge. A positive Lyapunov exponent corresponds to divergence between nearby geodesics, while a negative one signifies their convergence \cite{Cardoso2009, Pradhan2016, Modal2020}.

 Although the spacetime considered here has been reported previously \cite{rahaman}, only certain aspects of the optical properties of this spacetime have been explored. A systematic investigation encompassing strong gravitational lensing, timelike geodesic structure, strong-field observables and electromagnetic QNMs is still lacking, to the best of our knowledge. In the present work, we provide a unified treatment of these phenomena. In our previous work \cite{Roshila}, we investigated a black hole in Einstein–bumblebee gravity with a global monopole and cosmological constant, where the main focus was on field perturbations, greybody factors, absorption properties, and shadow characteristics. Although the present study is also concerned with black hole spacetimes containing monopole structures, the geometry considered here is different and includes an additional coupling parameter $\lambda$, leading to distinct physical behaviour. In particular, we examine how the global monopole parameter modifies the spacetime structure, influences observables associated with gravitational lensing,  geodesic motion and stability around the black hole, to further understand the gravitational aspects of the modified spacetime. 
 Furthermore, the quasinormal modes associated with electromagnetic perturbation of the modified spacetime will be calculated, illustrating the interplay between perturbative dynamics and geometry. The evolution of the electromagnetic perturbations in the time domain profile will also be studied. We also revisit shadow observables in order to determine the role of the parameter $\lambda$ and to obtain observational constraints from the EHT measurements of M87* and Sgr A*.  

The paper is structured as follows: In Section 2, we briefly review the spacetime under consideration and its horizon structure. We also investigate the behaviour of  the deflection angle and the impact on the lensing observables by the black hole parameters $\lambda$ and $\eta$. In Section 3, the timelike geodesic equation is derived and the general circular orbits and the ISCOs are analyzed using the derived effective potential. The stability of the timelike particle is also investigated using the Lyapunov exponent. Electromagnetic perturbation is discussed in Section 4, and the associated quasinormal mode frequencies are analyzed using the WKB-Pad\'{e} approximation and the improved AIM method in Section 5. In addition, the evolution of the electromagnetic perturbations in the time domain profile is investigated in Section 6. Finally, Section 7 summarizes the findings.      

\section{Strong lensing of a black hole with global monopole }\label{sec 2}
The standard global monopole model, comprising a self-coupled scalar triplet, is defined by the Lagrangian \cite{barriola, rahaman}
\begin{align}\label{lagrangian}
L=\frac{1}{2}\partial_\mu \phi^a \partial^\mu \phi^a - \frac{1}{4}\lambda (\phi^a \phi^a -\eta^2)^2,
\end{align}
where $\phi^a (a=1,2,3)$ and $\eta$ are self coupling triad of scalar fields and symmetry breaking scale respectively with arbitrary constant  $\lambda$  which measures the strength of the self interaction of the scalar field. \\
The model has an initial O(3) symmetry that undergoes spontaneous breaking to U(1). The corresponding monopole field configuration is given by
\begin{align}\label{monofield}
\phi^a=\eta \left(\frac{x^a}{r}\right) f(r),
\end{align} 
where the term $x^a=(r \sin\theta \cos\phi, r \sin\theta \sin\phi, r \cos\phi)$ such that $x^a x^a=r^2$.\\
Now, the  static, spherically symmetric spacetimes in polar coordinates can be written as
\begin{align}\label{metric}
ds^2= e ^{\nu(r)}dt^2 -e ^{\mu(r)}dr^2-r^2 d\Omega^2, 
\end{align}
where $d\Omega^2=d\theta^2 +   \sin^2\theta d\phi^2.$
For this spacetime background, the behavior of $f(r)$ in Eq. \eqref{monofield} is determined by the $\phi^a$ scalar field equation in the monopole ansatz
\begin{align}
e^{-\mu}f^{''}+e^{-\mu} \left[\frac{2}{r}+\frac{1}{2}(\nu^{'} +\mu^{'})\right]f^{'}-\frac{2f}{r^2}-\lambda \eta^2 f(f^2-1)=0,
\end{align}
where a primes  represents differentiation with respect to $r$. The size of the monopole core in flat space is $\delta\sim \lambda^{\frac{-1}{2}}\eta^{-1}$ and its mass is $M_{core}\sim \lambda^{\frac{-1}{2}}\eta$. If $\eta \ll m_p$, where $m_p$ represents the Planck's mass, we can assume that the structure of the monopole at small distance is not changed by gravity. Both the radius of the monopole core and its mass decrease with the increase of $\lambda$ and vice versa.\\
Ignoring all powers of $\frac{1}{r^2}$, Barriola-Vilenkin \cite{barriola} assumed $f=1$ outside the monopole core for flat space. To obtain an exact solution, Harari and Luostó \cite{harrari} solved the scalar field equation for flat space and yields as
\begin{align}\label{fieldequation}
f(r)=1-\frac{1}{x^2}-\frac{\frac{3}{2}-\tilde{\Delta}}{x^4}+O(x^{-6}).
\end{align}
where $x=\sqrt{\lambda} \eta r$ and $\tilde{\Delta}=8\pi G \eta^2$. From Eq. \eqref{fieldequation}, the stress energy components are obtained as \cite{rahaman}
\begin{align}
T^t_t=T^r_r=\frac{\eta^2}{r^2}-\frac{\lambda}{r^4}; T^{\theta}_{\theta}=T^{\phi}_{\phi}=\frac{\lambda}{r^4}.
\end{align}
The general solutions for the Einstein field equations are \cite{rahaman, FRahaman2025}
\begin{align}\label{gensoln}
e^{\nu}=e^{-\mu}=1-8\pi G \eta^2+\frac{8\pi G}{\lambda r^2}-\frac{M}{r},
\end{align}
where $M$ indicates the mass of the monopole core.
Eq. \eqref{gensoln} describes the gravitational field external to a mass distribution centred on a global monopole. Although the metric function in Eq. \eqref{gensoln} resembles the Reissner--Nordström spacetime because of the inverse-square term, its physical origin is fundamentally different. In the present case, the term $\frac{8\pi G}{\lambda r^2}$ arises from the scalar field configuration of the global monopole and is governed by the self-coupling constant $\lambda$, rather than by an electric charge. Following Ref. \cite{rahaman}, the self-coupling constant $\lambda$ is treated as a free model parameter, since no preferred numerical range or observational constraint is specified. The values of $\lambda$ adopted in this work are chosen to ensure physically admissible black hole solutions with real positive horizons and to illustrate the dependence of the physical observables on the self-coupling constant. A key application suggests that galaxies form when molecular clouds collapse gravitationally around such a monopole to create stars \cite{pando}. Depending on the initial collapse conditions, the star that forms may eventually becomes a black hole containing a global monopole.  Black hole configurations with global monopoles are theoretically motivated models for exploring how topological defects can influence black hole spacetimes and their observable properties. 

\begin{figure}[h!]
\centering
  \subfloat[\centering ]{{\includegraphics[width=170pt,height=170pt]{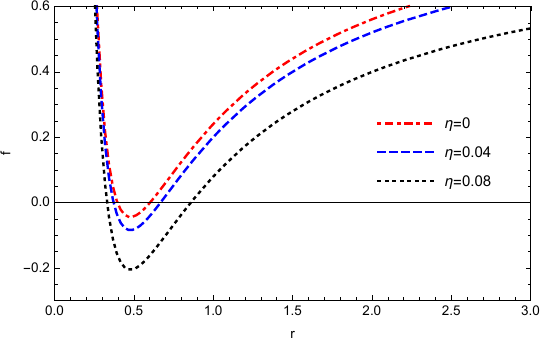}}\label{fig Freta}}
  \qquad
   \subfloat[\centering ]{{\includegraphics[width=170pt,height=170pt]{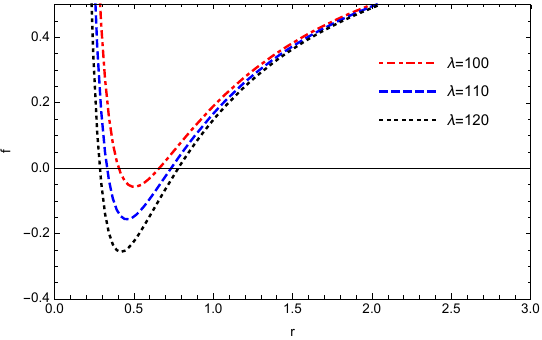}}\label{fig Frlamb}}
   \caption{Variation of $f(r)$ as a function of  $\eta$ (left) and $\lambda$ (right). Here $G=1$, $M=1$, with  $\lambda=105$ (left) and  $\eta=0.05$ (right).}
   \label{fig Fr}
\end{figure}

 The  necessary condition for the black hole to admit two positive real horizons is derived as
\begin{eqnarray}\label{eqn realroots}
M^2 \lambda > 32 \pi G(1-8 \pi G \eta^2).
\end{eqnarray} 
When $M^2 \lambda < 32 \pi G(1-8 \pi G \eta^2)$, $f(r)$ has no real root and the metric characterizes a naked singularity.   The presence of the  $\lambda$ dependent term $\frac{8\pi G}{\lambda r^2}$ modifies the radial behaviour of the metric function and consequently alters the horizon properties of the spacetime. Consequently, the spacetime admits a second horizon in addition to the event horizon, when the condition in Eq. \eqref{eqn realroots} is satisfied.
By solving equation $e^{\nu}=e^{-\mu}=0$, the two horizons of the black hole are obtained as 
\begin{eqnarray}
r_\pm = \dfrac{1}{2(1-k G \eta^2)}\left( M\pm \sqrt{M^2-\dfrac{4 k G}{\lambda}(1-k G \eta^2)}\right), 
\end{eqnarray} where $k=8 \pi$. The event horizon corresponding to $r=r_h$ is found to be
\begin{align}
r_h=\frac{M \lambda+\sqrt{\lambda} \sqrt{-32 G \pi+256 G^2 \pi^2 \eta^2+M^2 \lambda}}{2(\lambda-8G \pi \eta^2 \lambda)}.
\end{align}
In Fig. \ref{fig Fr}, we illustrate the behaviour of the metric function in terms of the radial coordinate $r$. As evident in the plots, we impose the condition given in Eq. \eqref{eqn realroots}, which gives two distinct horizons, the Cauchy horizon and the event horizon. We find that the distance between the horizons increases as the monopole parameter $\eta$ and the constant $\lambda$ increase. We also notice that the event horizon depends on changes in  $\eta$.

\begin{figure*}[!htbp]
\centering
\subfloat[\centering]
{\includegraphics[width=175pt,height=155pt]{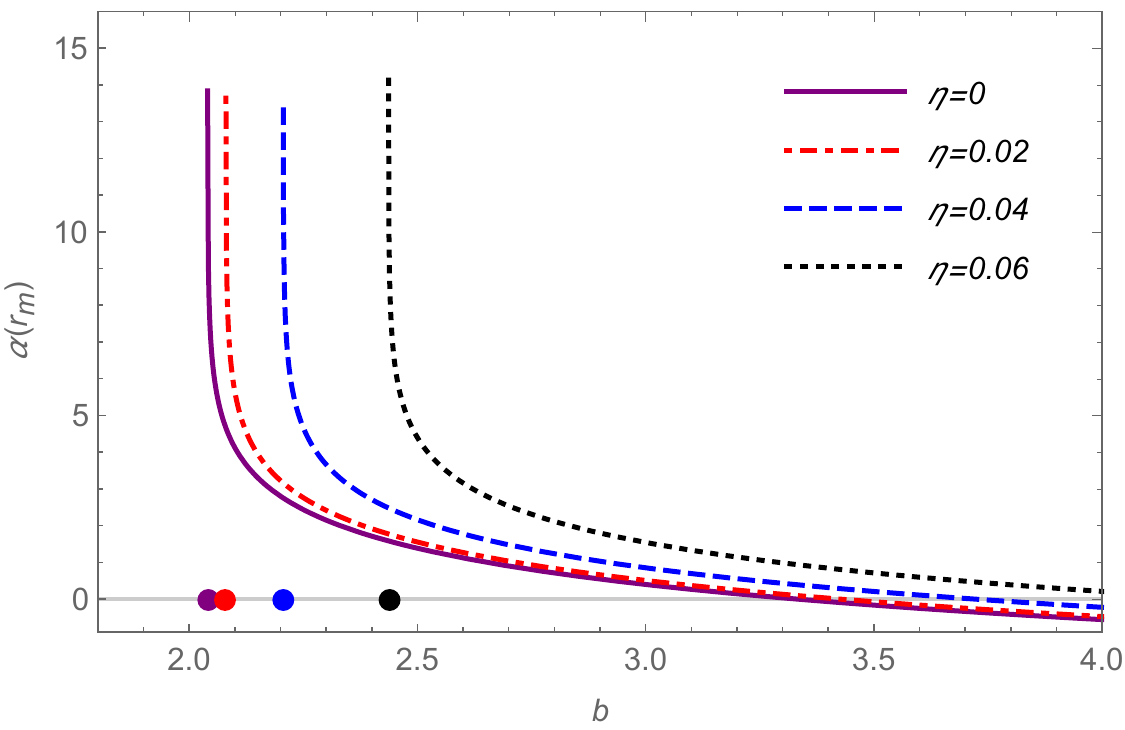}
\label {DAE}
}
\hfill
\subfloat[\centering]
{\includegraphics[width=175pt,height=155pt]{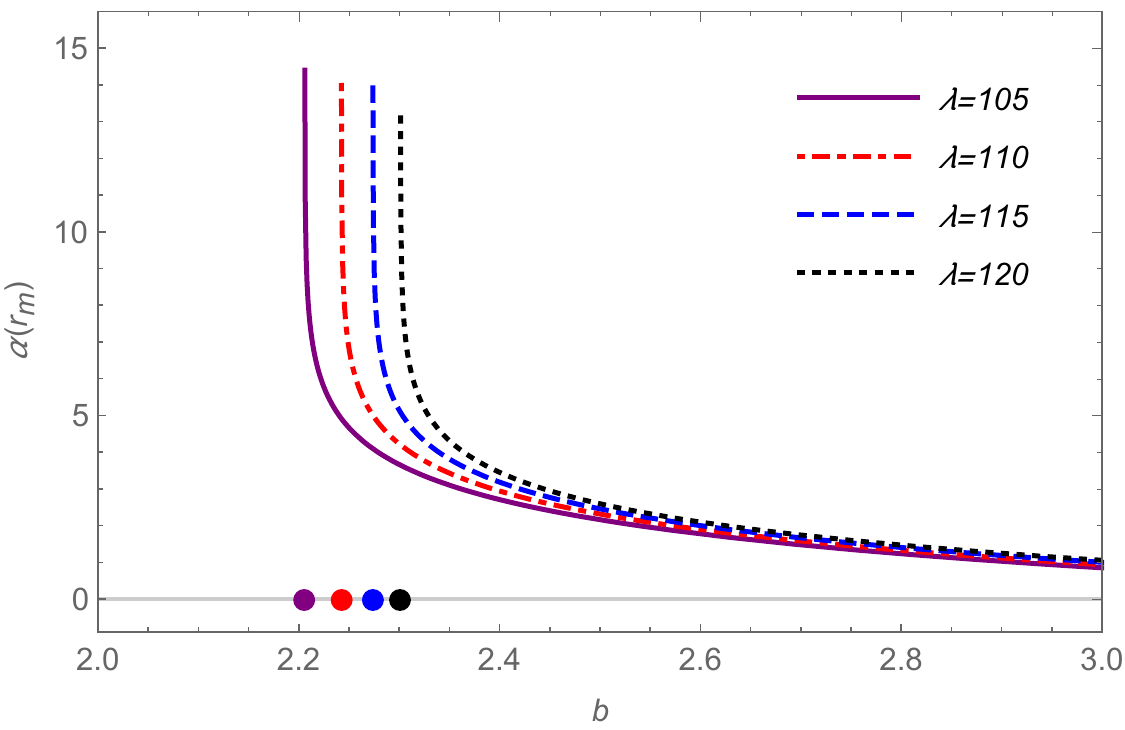}
\label {DAL}
}
\caption{The plot shows  $\alpha(r_m)$ changes with $b$ for various $\eta$ (left panel) and $\lambda$ (right panel). The colored points on the horizontal axis indicate $b_c$, where the deflection angle diverges. Here $G=1$, $M=1$, with  $\lambda=105$ (left) and  $\eta=0.04$ (right).}
\label{def}
\end{figure*}

The Lagrangian describing photon-orbit null geodesics around a black hole is
\begin{align}
\frac{1}{2} g_{\mu \nu} \dot{x}^\mu \dot{x}^\nu=0.
\end{align}\label{Lagrangian}
Here $\dot{x}^\mu$ corresponds to the wave number of the light ray, the dot denotes the differentiation with respect to an affine parameter. The angular momentum  $\mathcal{L}$ and the total energy $\mathcal{E}$ take the form
\begin{align}
&\mathcal{E}\equiv -g_{\mu \nu} t^\mu k^\nu=e^{\nu} \dot{t}, \hspace{0.5cm} \mathcal{L} \equiv g_{\mu \nu} \phi^\mu k^\nu=r^2 \dot{\phi}.
\end{align}
The impact parameter $b$ when $\mathcal{E}>0$ is defined by
\begin{align}
b\equiv \frac{\mathcal{L}}{\mathcal{E}}=\frac{r^2 \dot{\phi}}{e^{\nu} \dot{t}}.
\end{align}
Taking $\phi=\pi/2$, the trajectory of photon in the  spacetime of a black hole with a global monopole can be written as 
\begin{align}\label{trajectory}
e^{\nu}\dot{t}^2-e^{\mu}\dot{r}^2-r^2\dot{\phi}^2=0. 
\end{align}
Here, $\dot{r}^2=V_{eff}(r)$ and the effective potential $V_{eff}(r)$, which defines the possible photon orbits around the black hole is given by
\begin{align}\label{effpot}
V_{eff}(r)=\mathcal{E}^2-\frac{e^{\nu(r)}}{r^2}\mathcal{L}^2.
\end{align}
A photon travelling from a source toward an observer experiences gravitational deflection from the black hole at a distance $r_0$. A photon can only orbit if its effective potential $V_{eff}(r)$ is non-negative. Since $V_{eff}(r) \rightarrow E^2>0$ as $r\rightarrow \infty$, the photon can reach spatial infinity. The radius $r_m$ of the unstable circular photon orbit can be determined by applying the conditions: $V'_{eff}(r)=0$ and $V^{''}_{eff}(r)<0$. Therefore, the photon sphere radius $r_m$ corresponds to the greatest real solution of the equation
\begin{align}
\frac{g^{'}_{\theta\theta}(r)}{g_{\theta\theta}(r)}=\frac{g^{'}_{tt}(r)}{g_{tt}(r)}.
\end{align}
Solving the above condition yields
\begin{align}\label{rm}
r_m=\frac{3M \lambda+\sqrt{\lambda} \sqrt{-256 G \pi+2048 G^2 \pi^2 \eta^2+9 M^2 \lambda}}{4(\lambda-8G \pi \eta^2 \lambda)}.
\end{align} 
The impact parameter $b$ represents the perpendicular distance from the black hole's center to the photon's original trajectory. It influences the radial motion and determines the photon's minimum approach distance $r_0$. At this closest distance $r_0$, the impact parameter $b$ is expressed as
\begin{align}
b(r_0)=\sqrt{\frac{r^2_0}{e^{\nu(r_0)}}}.
\end{align}
The critical impact parameter $b_c$ when the closest approach distance $r_0 \rightarrow r_m$ is derived as \cite{bozza2002}
\begin{align}
b_c(r_m)&=\frac{r_m^2 \sqrt{2\lambda}}{\sqrt{M r_m \lambda-16\pi G}}.
\end{align}
In the regime of strong deflection, as $r_0 \rightarrow r_m$ or $b_0 \rightarrow b_c$, the impact parameter $b(r_0)$ may be expressed as a power series expansion in terms of $(r_0 - r_m)$ as
\begin{align}\label{bro}
b(r_0)&=b_c(r_m)+\frac{\sqrt{\lambda}(3M r_m \lambda-64\pi G)}{r_m^2 \sqrt{2}(M r_m \lambda-16\pi G)^{\frac{3}{2}}} (r_0-r_m)^2 +O(r_0-r_m)^3.
\end{align}
 The trajectory of Eq. \eqref{trajectory} can be rewritten as
\begin{eqnarray}
\Big(\frac{dr}{d\phi} \Big)^2&=A_0,
\end{eqnarray}
where the term $A_0$ is given by
\begin{align}
A_0&=\frac{8\pi G}{\lambda}\Big(\frac{r}{r_m}\Big)^4-\frac{M r^4}{r_m^3}+\frac{(1-8\pi G \eta^2)r^4}{r_m^2} -(1-8\pi G \eta^2)r^2+M r-\frac{8\pi G}{\lambda}. 
\end{align}
The deflection angle $\alpha(r_0)$ for the light ray is expressed by the following relation
\begin{align}
\alpha(r_0)=I(r_0)-\pi,
\end{align}
where 
\begin{align}\label{deangle}
I(r_0)\equiv \int_{r_0}^{\infty} \frac{2~dr}{\sqrt{A_0}}.
\end{align}
We may define a new variable $y$ as \cite{Tsukamoto2017}
\begin{align}
y \equiv 1-\frac{r_0}{r}.
\end{align}
Then Eq. \eqref{deangle} becomes
\begin{align}
I(r_0)=\int_{0}^{1} f(y,r_0) dy,
\end{align}
where
\begin{align}
f(y,r_0)=\frac{2r_0}{\sqrt{c_4(r_0)y^4+c_3(r_0)y^3+c_2(r_0)y^2+c_1(r_0)y}}.
\end{align}
Here, the  coefficients $c_n(r_0)$, $n=1,2,3,4$, are expressed as follows 
\begin{align}
&c_1(r_0)= 2r_0^2(1-8\pi G \eta^2)+\frac{32 \pi G}{\lambda}-3M r_0,\cr
&c_2(r_0)=-r_0^2(1-8\pi G \eta^2)-\frac{48 \pi G}{\lambda}+3M r_0 , \cr
&c_3(r_0)= \frac{32\pi G}{\lambda}-M r_0 ,\,\,\,
c_4=-\frac{8\pi G}{\lambda}.
\end{align}
In the regime of strong deflection,  $c_1(r_m)\rightarrow 0$ and  
\begin{align}
c_2(r_m)=r_m^2(1-8\pi G \eta^2)-\frac{16 \pi G}{\lambda}.
\end{align}
It is noted that $f(y,r_0)$ exhibits a divergence of order $y^{-1}$.\\
The term $I(r_0)$ can be expressed as:
\begin{align}
I(r_0)=I_R(r_0)+I_D(r_0).
\end{align}
The divergent part $I_D(r_0)$ is given by
\begin{align}\label{I1}
I_D(r_0)=\int_{0}^{1} f_D(y,r_0) dy,
\end{align}
where
\begin{align}
f_D(y,r_0)=\frac{2r_0}{\sqrt{c_2(r_0)y^2+c_1(r_0)y}}.
\end{align}
Then Eq. \eqref{I1} becomes
\begin{align}
&I_D(r_0)=\frac{4r_0}{\sqrt{c_2(r_0)}} \log \left[\frac{\sqrt{c_1(r_0)+c_2(r_0)}+\sqrt{c_2(r_0)}}{\sqrt{c_1(r_0)}}\right]. \nonumber\\
\end{align}
By using Eq. \eqref{bro}, $I_D(r_0)$ can be expressed in the strong deflection limit $r_0 \to r_m$ or $b \to b_c$  as
\begin{align}
I_D(r_m)=&-\bar{a} \log \Big(\frac{b}{b_c}-1 \Big)+\bar{a} \log  X +O((b-b_c)\times \log(b-b_c)),
\end{align}
where
\begin{align}
\bar{a}&=\frac{r_m\sqrt{2\lambda}}{\sqrt{3M r_m \lambda-64\pi G}}, 
X=\frac{2(3M r_m \lambda-64\pi G)}{M r_m \lambda-16\pi G}.
\end{align}
The regular part $I_R(r_0)$ is given by
\begin{align}
I_R(r_0)\equiv \int_{0}^{1} f_R(y,r_0) dy,
\end{align}
where
\begin{align}
f_R(y,r_0)\equiv f(y,r_0)-f_D(y,r_0).
\end{align}
We consider
\begin{align}
 \lim_{r_0 \rightarrow r_m}f_R(y,r_0)&= \frac{2r_m}{y\sqrt{c_4(r_m) y^2+c_3(r_m) y+c_2(r_m)}}-\frac{2r_m}{y\sqrt{c_2(r_m)}} .
\end{align}
An analytical expression in the strong deflection limit $r_0 \rightarrow r_m$ or $b_0 \rightarrow b_c$ can be obtained as
\begin{align}
I_R(r_m)&= \bar{a} \,\, \log \Big[ \frac{4(3M r_m \lambda-64\pi G)^2}{M^2 r_m^2 \lambda^2(M r_m \lambda-16\pi G)} \times \cr & \Big(2\sqrt{M r_m \lambda-16\pi G}-\sqrt{3M r_m \lambda-64\pi G)^2} \Big)^2 \Big] +O((b-b_c)\log(b-b_c)).
\end{align}
Thus, the deflection angle $\alpha(r_m)$ is given by
\begin{align}\label{defangle}
\alpha(r_m)=-\bar{a}\,\, \log \Big(\frac{b}{b_c}-1 \Big)+\bar{b}+O((b-b_c)\log(b-b_c)),
\end{align}
where
\begin{align}
\bar{b}=&\bar{a}\,\, \log \Big[\frac{8(3M r_m \lambda-64\pi G)^3}{M^2 r_m^2 \lambda^2(M r_m \lambda-16\pi G)^2} \times \Big(2\sqrt{M r_m \lambda-16\pi G}-\sqrt{3M r_m \lambda-64\pi G)^2} \Big)^2 \Big] \cr &-\pi.\nonumber\\
\end{align}
Here, $\bar{a}$ and $\bar{b}$ represent the lensing coefficients. We discuss the behavior of the deflection angle $\alpha(r_m)$ by the black hole parameters $\eta$ and $\lambda$ in Fig. \ref{def}. It is observed that the deflection angle influence by global monopole diverges at larger values of $b_c$, $\eta$ and $\lambda$. This finding indicates that the deflection angle is directly dependent on $\eta$ and $\lambda$, which rise as $\eta$ and $\lambda$ increase.

\subsection{Lens observation}
In this section, we analyze the strong-field lensing observables associated with the black hole. Since the spacetime containing a global monopole is not asymptotically flat but instead possesses an asymptotically conical geometry characterized by the deficit factor $1-8\pi G\eta^2$, the optical geometry differs from that of the Schwarzschild spacetime. In the present work, we follow the conventional strong-deflection framework introduced by Bozza and subsequently adopted for black holes with global monopoles. Within this framework, the asymptotic deficit is incorporated through the metric functions entering the null geodesic equations and the corresponding impact parameter, while the standard strong-deflection lens equation is employed to evaluate the observable quantities. The resulting observables should therefore be interpreted within this established strong-field framework. Accordingly, the source and the observer are assumed to be located in the asymptotic region of the global-monopole spacetime, sufficiently far from the lens and nearly aligned with each other \cite{bozza2002,bozza2008,PRD2015,PRD2025}. The corresponding lens equation is given by
\begin{align}\label{lenseqn}
\beta=\theta-\frac{D_{LS}}{D_{OS}} \Delta \alpha_n,
\end{align}
where $\theta$ represents the angular distance between the lens and the image, and $\beta$ denotes the angular position of the source. $D_{OL}$ denotes the distance between the observer and the lens, and  $D_{LS}$ denotes the distance between the lens and the source.\\
Using Eqs. \eqref{defangle} and \eqref{lenseqn} and the relation $b\approx \theta D_{OL}$, the position of $n^{th}$ relativistic image $\Delta \alpha_n$ can be approximate as \cite{Bozza2003, Y. Lan 2025}
\begin{align}\label{thetan}
\theta_n \simeq \theta_n^0+\frac{D_{OS}}{D_{LS}} \frac{b_c e_n}{D_{OL} \bar{a}} (\beta-\theta_n^0),
\end{align}
where
\begin{align}\label{l2}
\theta_n^0=\frac{b_c}{D_{OL}}(1+e_n), 
\end{align}
\begin{align}
e_n=e^{\frac{\bar{b}-2n\pi}{\bar{a}}}.
\end{align}

\begin{figure*}[!htbp]
\centering
\subfloat[\centering]
{\includegraphics[width=175pt,height=155pt]{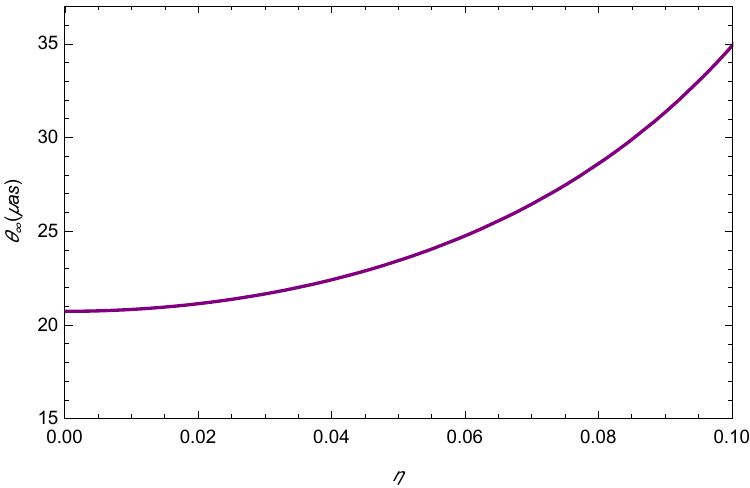}
\label {thetaeta}
}
\hfill
\subfloat[\centering]
{\includegraphics[width=175pt,height=155pt]{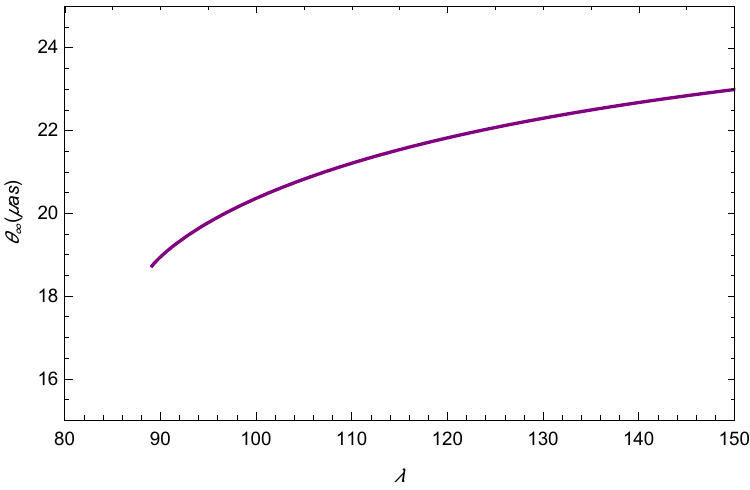}
\label {thetalambda}
}
\caption{Variation of the strong lensing observable $\theta_{\infty}$ in strong field limit with respect to the parameters $\eta$ (left) and $\lambda$ (right). Here $G=1$, $M=1$, with  $\lambda=105$ (left) and  $\eta=0.01$ (right).}
\label{thetainf}
\end{figure*}

\begin{figure*}[!htbp]
\centering
\subfloat[\centering]
{\includegraphics[width=175pt,height=155pt]{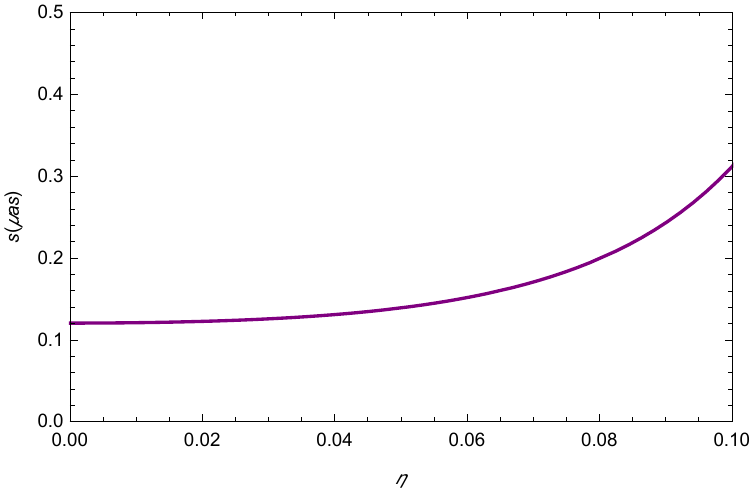}
\label {seta}
}
\hfill
\subfloat[\centering]
{\includegraphics[width=175pt,height=155pt]{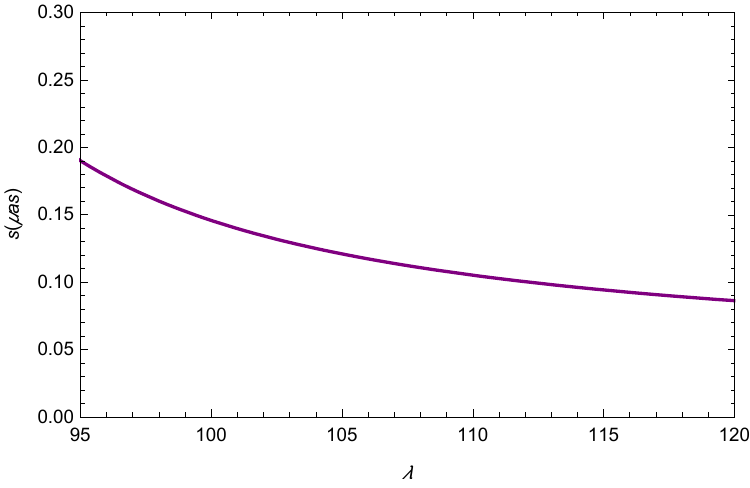}
\label {slambda}
}
\caption{Variation of the strong lensing observable $s$ in the strong field limit with respect to the parameters $\eta$ (left) and $\lambda$ (right). Here $G=1$, $M=1$, with  $\lambda=105$ (left) and  $\eta=0.01$ (right).}
\label{sep}
\end{figure*}

\begin{figure*}[!htbp]
\centering
\subfloat[\centering]
{\includegraphics[width=175pt,height=155pt]{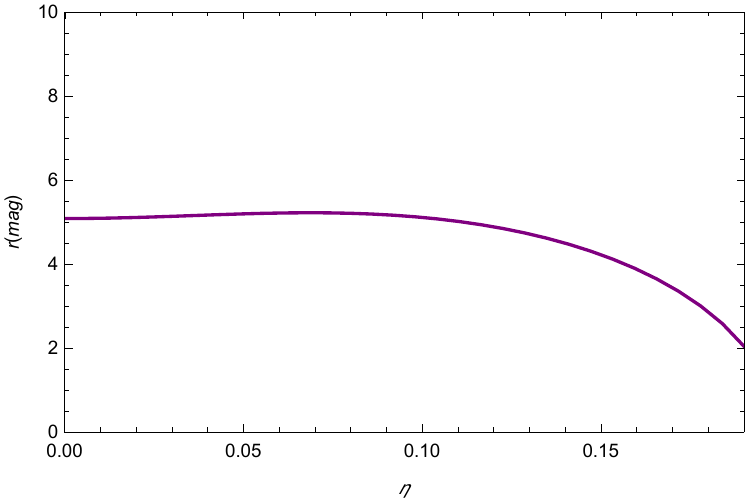}
\label {reta}
}
\hfill
\subfloat[\centering]
{\includegraphics[width=175pt,height=155pt]{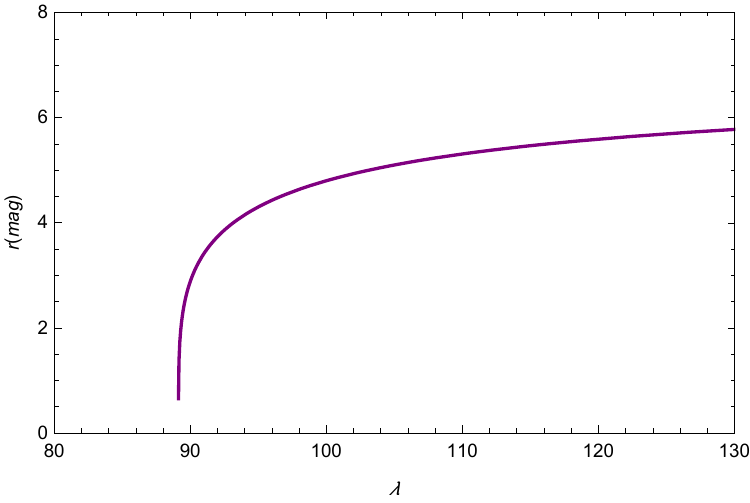}
\label {rlambda}
}
\caption{Variation of the strong lensing observable $r_{mag}$ in strong field limit with respect to the parameters $\eta$ (left) and $\lambda$ (right). Here $G=1$, $M=1$, with  $\lambda=105$ (left) and  $\eta=0.01$ (right).}
\label{Rmag}
\end{figure*}

In this expression, $\theta_n^0$ denotes the angular position of the image that corresponds to a photon having completed $2n\pi$ revolutions. Since surface brightness is conserved in gravitational lensing, the magnification equals the ratio of the solid angle subtended by the $n^{th}$ image to that of the source. For the $n^{th}$ relativistic image, the magnification is given by \cite{bozza2002}
\begin{align}\label{relat}
\mu_n=\Big(\frac{\beta}{\theta} \frac{d\beta}{d\theta} \Big|_{\theta_n^0} \Big)^{-1}=e_n \frac{b_c^2  D_{OS}(1+e_n)}{\bar{a} \beta D_{OL}^2 D_{LS}}.
\end{align}
In the case of perfect source alignment $(\beta=0)$, the above Eq. \eqref{relat} diverges. This divergence corresponds to the maximum probability of detecting a gravitationally lensed image. As the magnification is in inverse proportion to $D_{OL}^2$, all  images are inherently dim and their brightness decreases  as $n$ increases. Consequently, higher-order images become progressively less visible, making the brightness of the first relativistic image ($\theta_1$) dominant. This leads to a simplified observational picture: $\theta_1$ appears as a distinct outermost image, while the remaining higher-order images are packed together near $\theta_\infty$.
Therefore, we define three essential observables as \cite{bozza2002}
\begin{align}\label{three}
& \theta_{\infty}=\frac{b_c}{D_{OL}}, \cr
& s=\theta_{\infty} e^{\frac{\bar{b}-2\pi}{\bar{a}}}, \cr
& r=e^{\frac{2\pi}{\bar{a}}},\,\,\,
r_{\text{ mag}}= \frac{5\pi}{\bar{a}\,\, \text{ ln} 10}.
\end{align}
From the above Eq. \eqref{three}, $s$ represents the angular separation between $\theta_1$ and $\theta_\infty$, $r_{\text{ mag}}$ denotes the ratio between the flux of the first image and the combined flux of all the remaining images. It is noted that $r_{\text{ mag}}$ does not depend on $D_{OL}$. The behavior of the lensing observables $\theta_\infty$, $s$ and $r_{\text{ mag}}$ for varying $\eta$ and $\lambda$ are depicted in Figs. \ref{thetainf}, \ref{sep} and \ref{Rmag} respectively. We observe from the figures that $\theta_\infty$ increases for both increasing $\eta$ and $\lambda$, $s$ increases for increasing $\eta$ but it has an opposite effect for $\lambda$. For increasing $\eta$, the behavior of $r_{\text{ mag}}$ slowly increases initially and then decreases but $r_{\text{ mag}}$ increases with increasing $\lambda$. The lensing coefficient $\bar{a}$ decreases for small values of $\eta$ and then increases for higher values of $\eta$ while $\bar{a}$ decreases monotonically for increasing $\lambda$. The lensing coefficient $\bar{b}$ increases with higher values of $\eta$ and $\lambda$. The numerical values are also displayed in Tables \ref{tab lensing} and \ref{tab lensing2}.

\begin{table}[h]
 \centering
    \begin{tabular}{c| c c c c c c c }
    $\eta$   &   $\theta_{\infty}(\mu a s) $ &  $s(\mu a s)$ &  $r_m(mag)$ &$\bar{a}$ & $\bar{b}$ &  $\theta^E_1$ & $\Delta T_{2,1}^s$   \\  
\hline
0    & 20.721 & 0.120374  &  5.09101 &  1.33999  & -0.615467  & 20.8414  & 12.8211  \\
0.02 & 21.1269 & 0.122481  &  5.11645  & 1.33332  & -0.583899  &  21.2494  &  13.0722  \\
0.04 & 22.4017 & 0.130763  &  5.17437  & 1.3184  & -0.498002  &  22.5325  & 13.861  \\ 
0.06 & 24.7442 & 0.151544  &  5.22253  & 1.30624  & -0.372725  &  24.8958  & 15.3104  \\
0.08 & 28.5916 & 0.199119  &  5.21634  & 1.30779  & -0.212569  &  28.7907  & 17.691  \\
0.1  & 34.8641 & 0.311307  &  5.11815  &  1.33288 & -0.00591895  &  35.1754  & 21.5721   \\
 \end{tabular}
\caption{Lensing observables computed in the strong-field regime, along with lensing coefficients, as $\eta$ varies and $\lambda=105$ with $M=1$.}
    \label{tab lensing}
\end{table}

\begin{table}[h]
 \centering
    \begin{tabular}{c| c c c c c c c }
   $\lambda$  &   $\theta_{\infty}(\mu a s) $ &  $s(\mu a s)$ &  $r_m(mag)$ &$\bar{a}$ & $\bar{b}$ &  $\theta^E_1$ & $\Delta T_{2,1}^s$   \\  
\hline
105   & 20.8216  & 0.12087 & 5.09769 & 1.33823  & -0.607405 & 20.9425 & 12.8833 \\
110   & 21.2067  & 0.10506 & 5.30725 & 1.28539  & -0.539072 & 21.3117 & 13.1216 \\
115   & 21.5349  & 0.0941207 & 5.46496 & 1.24829 & -0.498613 & 21.629 & 13.3247  \\ 
120   & 21.8198  & 0.0860983 & 5.58936 & 1.22051 & -0.472458 & 21.9059 & 13.501 \\
125   & 22.0705  & 0.0799612 & 5.69068 &  1.19878 & -0.45451 & 22.1505 & 13.6561  \\
130   & 22.2934  & 0.0751133 & 5.77521 &  1.18124 & -0.441647 & 22.3685 & 13.794    \\
 \end{tabular}
\caption{Lensing observables computed in the strong-field regime, along with lensing coefficients, as $\lambda$ varies and $\eta=0.01$ with $M=1$.}
    \label{tab lensing2}
\end{table}

\begin{figure}[h!]
\centering
  \subfloat[\centering ]{{\includegraphics[width=170pt,height=170pt]{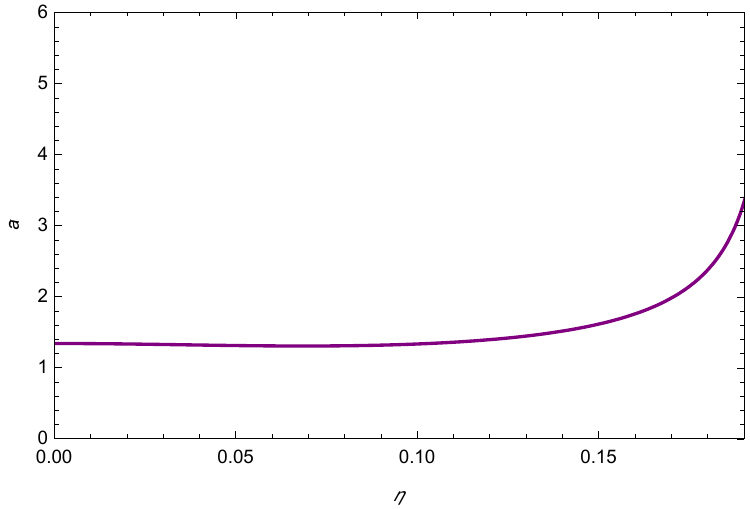}}\label{abarE}}
  \qquad
   \subfloat[\centering ]{{\includegraphics[width=170pt,height=170pt]{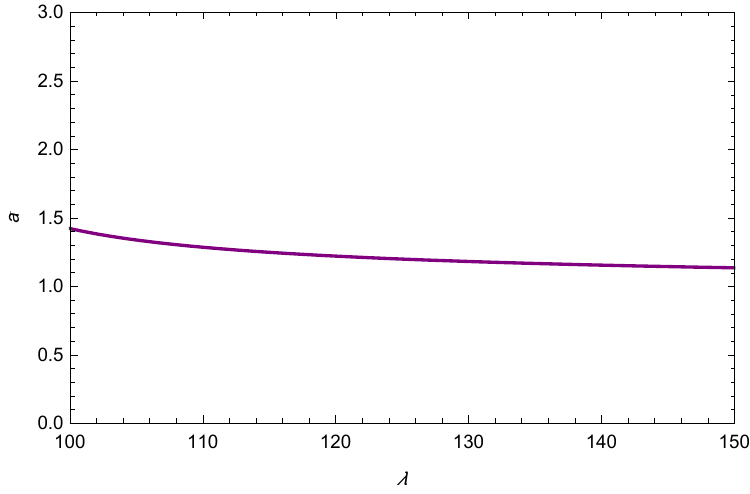}}\label{abarL}}
\caption{Lensing coefficient $\bar{a}$ in the strong field limit versus the parameters $\eta$ (left) and $\lambda$ (right). Here $G=1$, $M=1$, with  $\lambda=105$ (left) and  $\eta=0.01$ (right).}
\label{abar}
\end{figure}

\begin{figure}[h!]
\centering
  \subfloat[\centering ]{{\includegraphics[width=170pt,height=170pt]{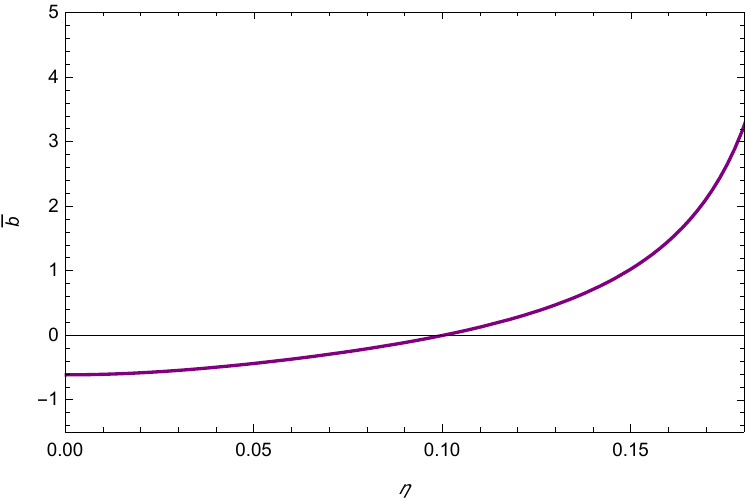}}\label{beta}}
  \qquad
   \subfloat[\centering ]{{\includegraphics[width=170pt,height=170pt]{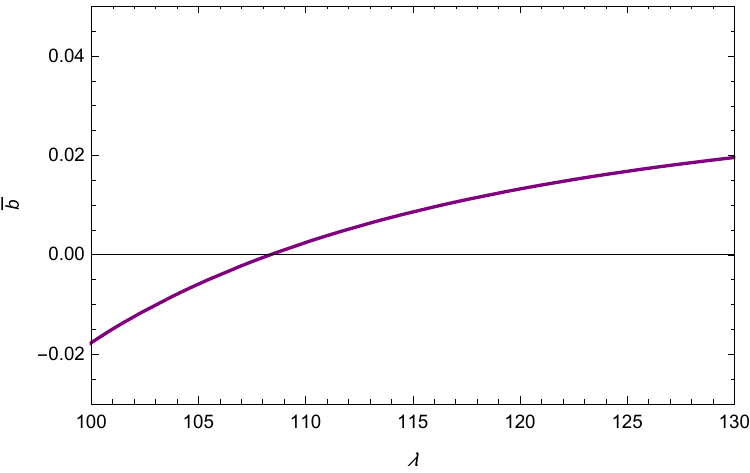}}\label{blambda}}
\caption{Lensing coefficient $\bar{b}$ in the strong field limit versus the parameters $\eta$ (left) and $\lambda$ (right). Here $G=1$, $M=1$, with  $\lambda=105$ (left) and  $\eta=0.01$ (right).}
\label{bbar}
\end{figure}

\subsection{Einstein ring}
\begin{figure}[h!]
\centering
  \subfloat[\centering ]{{\includegraphics[width=190pt,height=160pt]{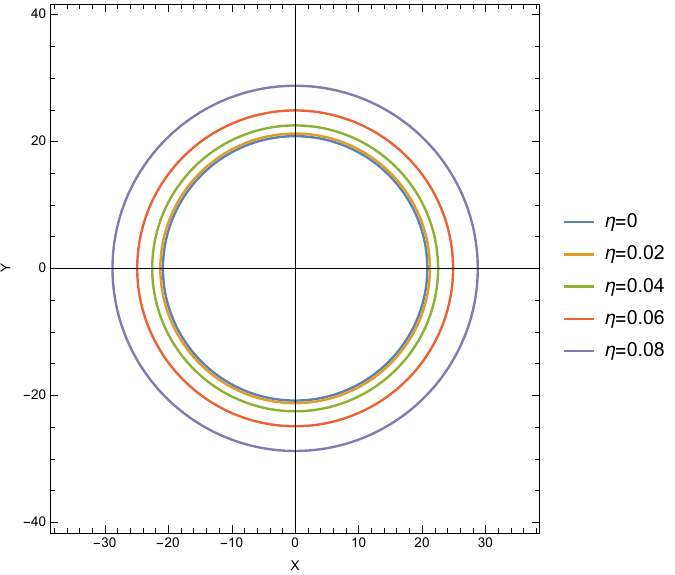}}\label{fig EringEta}}
  \qquad
   \subfloat[\centering ]{{\includegraphics[width=190pt,height=160pt]{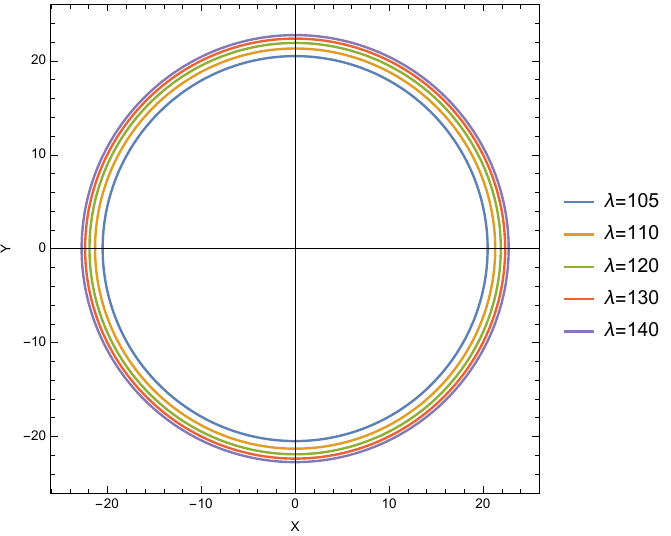}}\label{fig EringLambda}}
   \caption{Plot of the outermost Einstein radius for different values of  $\eta$ (left) and $\lambda$ (right). Here $G=1$, $M=1$, with  $\lambda=105$ (left) and  $\eta=0.01$ (right).}
   \label{fig Ering}
\end{figure}

\begin{figure*}[!htbp]
\centering
\subfloat[\centering]
{\includegraphics[width=175pt,height=155pt]{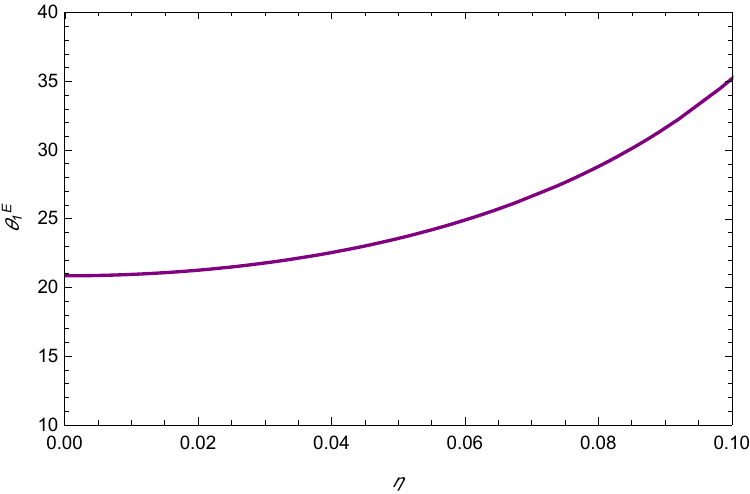}
\label {eeta}
}
\hfill
\subfloat[\centering]
{\includegraphics[width=175pt,height=155pt]{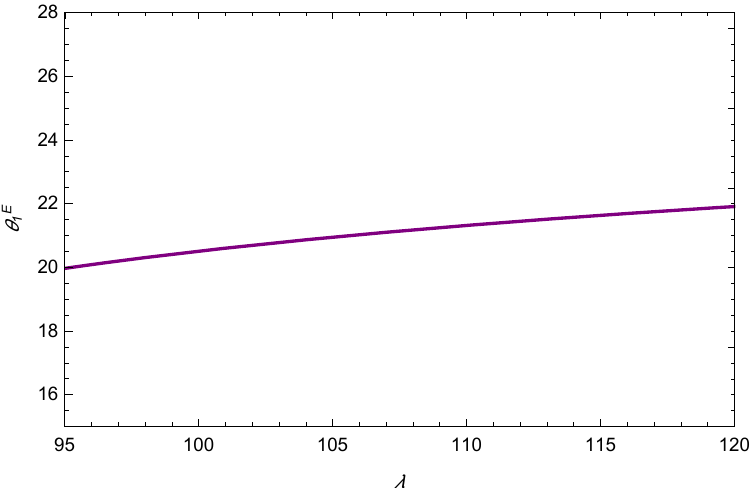}
\label {elambda}
}
\caption{Behavior of the Einstein ring $\theta_1^E$ in strong field limit with respect to the parameters $\eta$ (left) and $\lambda$ (right). Here $G=1$, $M=1$, with  $\lambda=105$ (left) and  $\eta=0.01$ (right).}
\label{einstein}
\end{figure*}

 When a background light source, the deflecting lens, and the observer are situated along a direct line of sight, an Einstein ring forms. This phenomenon is driven by the surrounding gravitational field, which warps spacetime and generates concentric relativistic images of the positioned source. However, if only one source point is precisely aligned, a complete relativistic Einstein ring can be formed \cite{Tsukamoto2017}. Thus, when the source, lens and observer are perfectly aligned $i.e. (\beta=0)$, Eq. \eqref{lenseqn} can be written as
\begin{align}\label{er1}
\theta^{E}_n=\left(1-\frac{D_{OL}}{D_{LS}}\frac{b_c e_n}{D_{OL} \bar{a}}\right)\theta^0_n .
\end{align}
For the case of perfect alignment, with the lens positioned at the midpoint between source and observer, the angular radius of the Einstein ring can be derived from Eqs. \eqref{l2} and \eqref{er1} as 
\begin{align}\label{ringsnew}
\theta^{E}_n=\left(1-\frac{2  e_n b_c }{ \bar{a} D_{OL} }\right) \times \frac{(1+e_n)b_c}{D_{OL}}.
\end{align}
Taking $D_{OS}=2D_{OL}$ and $D_{OL}\gg b_c$, the above equation reduces to
\begin{align}\label{er2}
\theta^{E}_n=\frac{(1+e_n) b_c}{D_{OL}}.
\end{align}
Equation \eqref{er2} defines the radius of the $n^{th}$ relativistic Einstein ring. In this case, the outermost ring occurs when $n=1$ and the radius of the ring gets smaller as $n$ increases.  The radius of the outermost Einstein ring increases with the rise of $\eta$ and $\lambda$ as shown in Fig. \ref{fig Ering}. In Fig. \ref{einstein}, both the graphs of $\theta^{E}_1$ are found increasing for varying $\eta$ and $\lambda$.

\subsection{Time delay}
\begin{figure*}[!htbp]
\centering
\subfloat[\centering]
{\includegraphics[width=175pt,height=155pt]{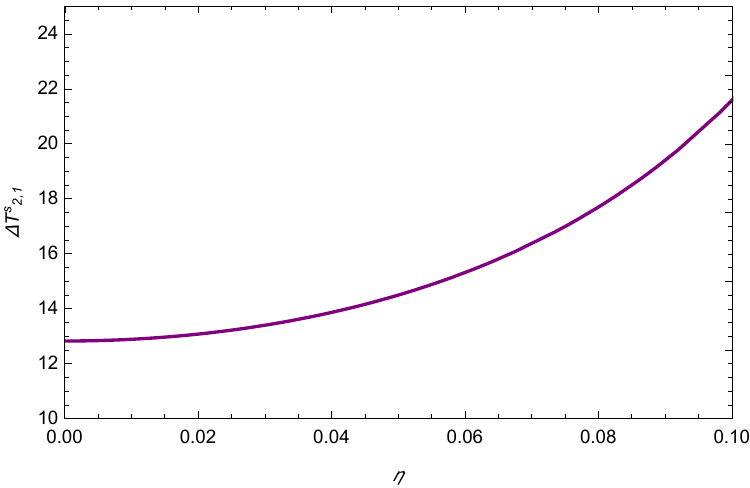}
\label {teta}
}
\hfill
\subfloat[\centering]
{\includegraphics[width=175pt,height=155pt]{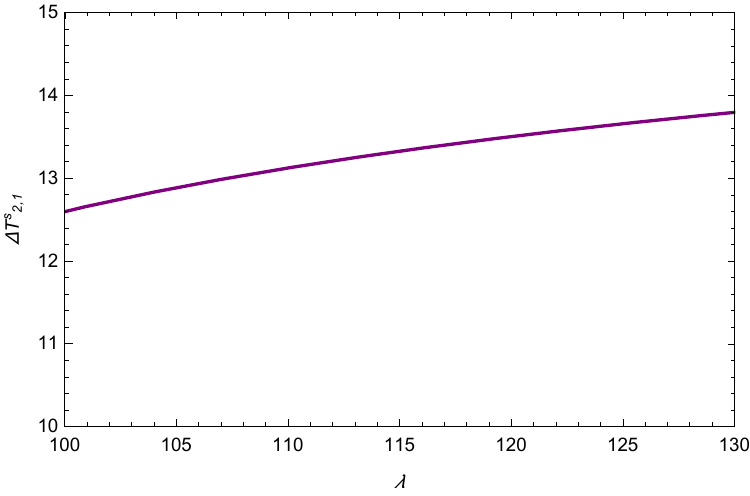}
\label {tlambda}
}
\caption{Behavior of the time delay $\Delta T^{s}_{2,1}$ in strong field limit with respect to the parameters $\eta$ (left) and $\lambda$ (right). Here $G=1$, $M=1$, with  $\lambda=105$ (left) and  $\eta=0.01$ (right).}
\label{timedelay}
\end{figure*}

 This subsection will analyse another key observable in strong gravitational lensing called the time delay. The light rays that form various relativistic images travel different paths, causing them to reach the observer at different times. The time delay arises from the discrepancy in formation time between two relativistic images. This discrepancy occurs because photons follow different trajectories around the black hole, leading to unequal travel times along the distinct paths associated with each image. Hence, there exists a time offset among the various relativistic images. From observed time signals of two such images, one can calculate the time delay. The duration a photon spends to complete an orbit around the black hole is given by \cite{molla}
\begin{align}\label{time delay 1}
T(\tilde{b})=\bar{a} \log \left(\dfrac{b}{b_c}-1\right) +\bar{b}+O\left(b-b_c\right).
\end{align}
When the first and second images lie on the same side of the lens, the time delay between them can be approximated by
\begin{align}
\Delta T^{s}_{2,1}=2 \pi b_c=2\pi D_{OL} \theta_\infty.
\end{align}
Equation \eqref{er2} suggests that the quantum behavior of black holes can be determined with the same level of accuracy if a precise time-delay measurement and a small inaccuracy in the crucial impact parameter are provided. In Fig. \ref{timedelay} and Tables \ref{tab lensing} and \ref{tab lensing2}, the time delay $\Delta T^{s}_{2,1}$ between the second and first relativistic images increases when both the parameters $\eta$ and $\lambda$ increase simultaneously.

\subsection{Shadow radius of black hole}
\begin{figure}[h!]
\centering
  \subfloat[\centering ]{{\includegraphics[width=170pt,height=170pt]{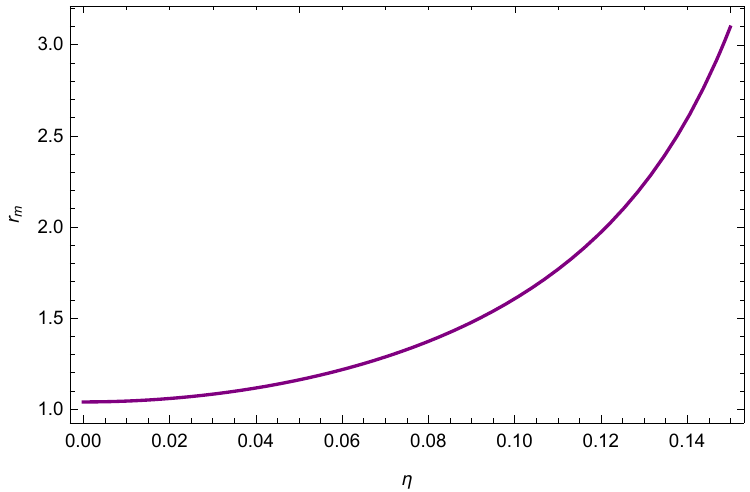}}\label{fig rpeta1}}
  \qquad
   \subfloat[\centering ]{{\includegraphics[width=200pt,height=170pt]{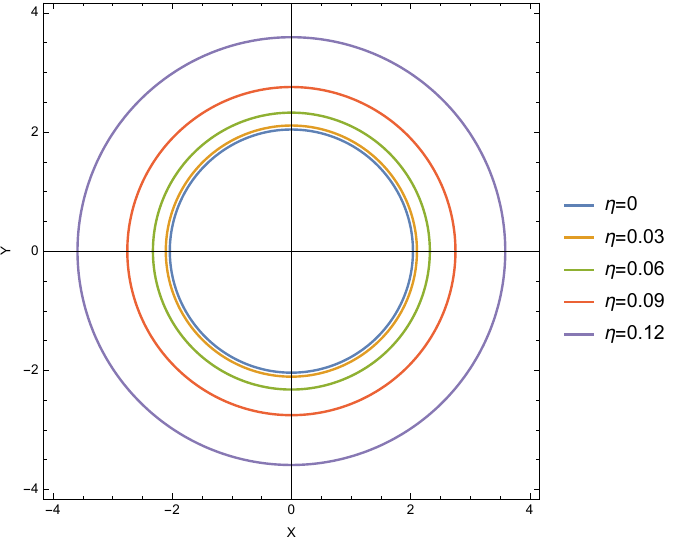}}\label{fig 2Shadoweta}}
   \caption{Plot of photon sphere (left) and the shadow radius (right) for varying  $\eta$. Here, $M=1$, $\lambda=105$, $\mathcal{L}=1$, $G=1$.}
   \label{fig Shadow1}
\end{figure}

\begin{figure}[h!]
\centering
  \subfloat[\centering ]{{\includegraphics[width=170pt,height=170pt]{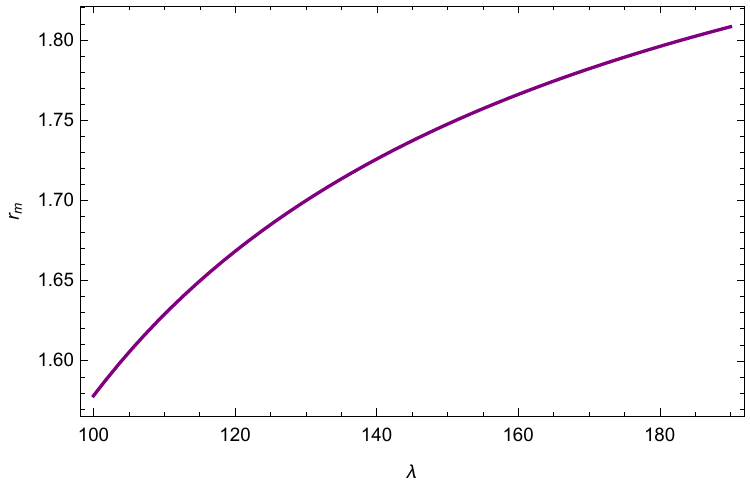}}\label{fig rplamb1}}
  \qquad
   \subfloat[\centering ]{{\includegraphics[width=200pt,height=170pt]{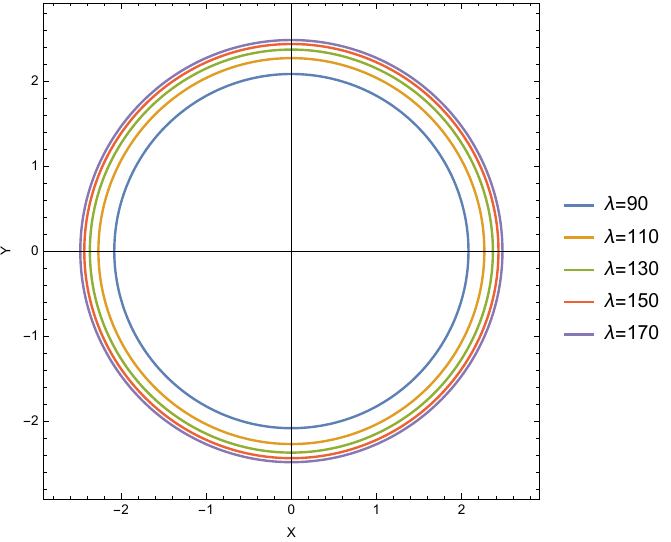}}\label{fig 2Shadowlamb}}
   \caption{Plot of the photon sphere (left) and the shadow radius (right) for different values of $\lambda$. Here, $M=1$, $\eta=0.1$, $\mathcal{L}=1$, $G=1$.}
   \label{fig Shadow2}
\end{figure}

The black hole's gravitational field deflects light coming from faraway celestial bodies. Some of the photons that move along particular unstable circular orbits are trapped by the gravitational field, creating a photon sphere. In this section, we examine the effect of the global monopole parameter $\eta$ and the parameter $\lambda$ on the radius of the black hole's shadow. For the equatorial circular motion,  the following conditions should be fulfilled by the null-like geodesics \cite{PriyoEPJC}

\begin{align}\label{eqn Vrp}
V(r)\vert_{r=r_m}=0 \hspace{0.5cm} \text{and} \hspace{0.5cm}  V'(r)\vert_{r=r_m}=0.
\end{align}
  For a distant static observer situated at $r_0 \to \infty$, the observed black hole shadow radius $\mathcal{R}_s$ is defined as \cite{Ahmad2025, Roshila} 
\begin{align}\label{eqn sha}
\mathcal{R}_s=r_m \sqrt{\dfrac{f(r_0)}{f(r_m)}} = \dfrac{r_m \sqrt{1-8 \pi G \eta^2}}{ \sqrt{f(r_m)}},
\end{align} 
where $f(r_0 \to \infty) = 1-8 \pi G \eta^2$. Here, the factor $\sqrt{1-8\pi G \eta^2}$ explicitly accounts for the asymptotic non-flatness of the spacetime, properly normalizing the coordinate time for an observer at infinity. In the equatorial plane $(\theta={\pi}/{2})$, the shadow radius of the black hole $\mathcal{R}_s$ corresponds to the photon sphere's critical impact parameter. 
In Figs. \ref{fig Shadow1} and \ref{fig Shadow2}, we illustrate the behaviour of the photon sphere $r_m$ and the shadow radius $R_s$ for various values of the parameters $\eta$ and $\lambda$, respectively. We notice from Fig. \ref{fig rpeta1} that the photon sphere radius is enlarged as $\eta$ grows and the rate of increase rises with $\eta$. Also, we see from Fig. \ref{fig rplamb1} that the photon sphere rises monotonically with $\lambda$ but the rate of increase gradually declines with $\lambda$. These observed behaviours of $r_m$ are reflected in the shadow radius since the shadow radius depends on $r_m$ as given in Eq. \eqref{eqn sha}.  From Figs. \ref{fig 2Shadoweta} and \ref{fig 2Shadowlamb}, we notice a perfectly circular shadow for varying $\eta$ and $\lambda$ as observed for a non rotating black hole, but the size of the shadow changes depending upon the spacetime parameters. Furthermore, It is observed from Fig. \ref{fig 2Shadoweta} that increasing the monopole parameter $\eta$ leads to increase in the shadow radius and the increase in the shadow size is more pronounced for larger values of $\eta$. This behaviour indicates that the global monopole parameter modifies the spacetime geometry in a way that enhances the effect of light bending near the black hole, which ultimately increases the size of the shadow. Moreover, increasing $\lambda$ also leads to a monotonic increase in the  black hole's shadow radius however the rate of increase progressively declines. Also, we notice that the shadow size is more sensitive to changes in the monopole parameter. These variations in the shadow radius closely follow the observed behaviour of the corresponding photon sphere radius. For different values of $\eta$ and $\lambda$, the photon radius and the shadow radius are shown in Table \ref{tab rpRs}. We can see that the shadow radius and the photon radius increase with increasing $\eta$ and $\lambda$, which coincide+ with the results shown in Figs. \ref{fig Shadow1} and \ref{fig  Shadow2}. The data in the table suggests that $\eta$ and $\lambda$ give substantial impact on the photon radius and black hole shadow.

\begin{table}[h]
 \centering
    \begin{tabular}{c| c c c | c c c }
    $\eta$   &  0 &   0.04 &  0.08 & 0.05 & 0.05 &  0.05  \\  
\hline
$\lambda$ &   105 &   105 &  105 & 95 &    110 & 130\\
$r_{p}$    &    1.03945 &   1.11586 &   1.3716  & 1.07573 &   1.19125  &  1.27764  \\  

$R_{s}$  &   2.04054  &  2.16124 &   2.57924   & 2.14282 &   2.26724   &   2.3676   \\

 \end{tabular}
\caption{The photon radius $r_p$ and the shadow radius $R_s$ for varying $\eta$ and $\lambda$ with $M=1$ and $G=1$.}
    \label{tab rpRs}
\end{table}

\begin{figure}[h!]
\centering
  \subfloat[\centering ]{{\includegraphics[width=170pt,height=170pt]{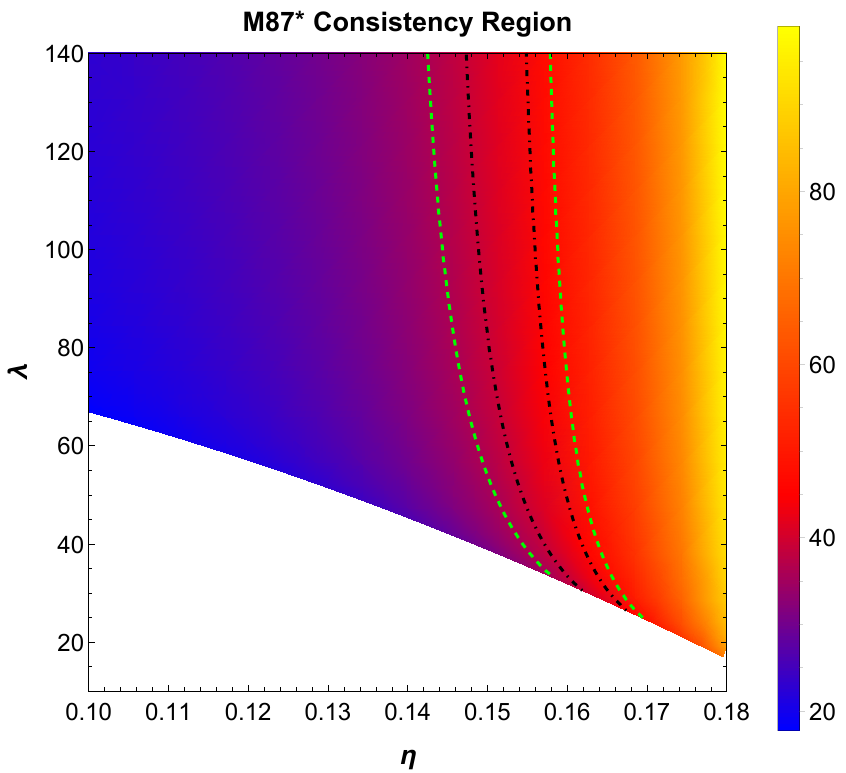}}\label{fig M87}}
  \qquad
   \subfloat[\centering ]{{\includegraphics[width=200pt,height=170pt]{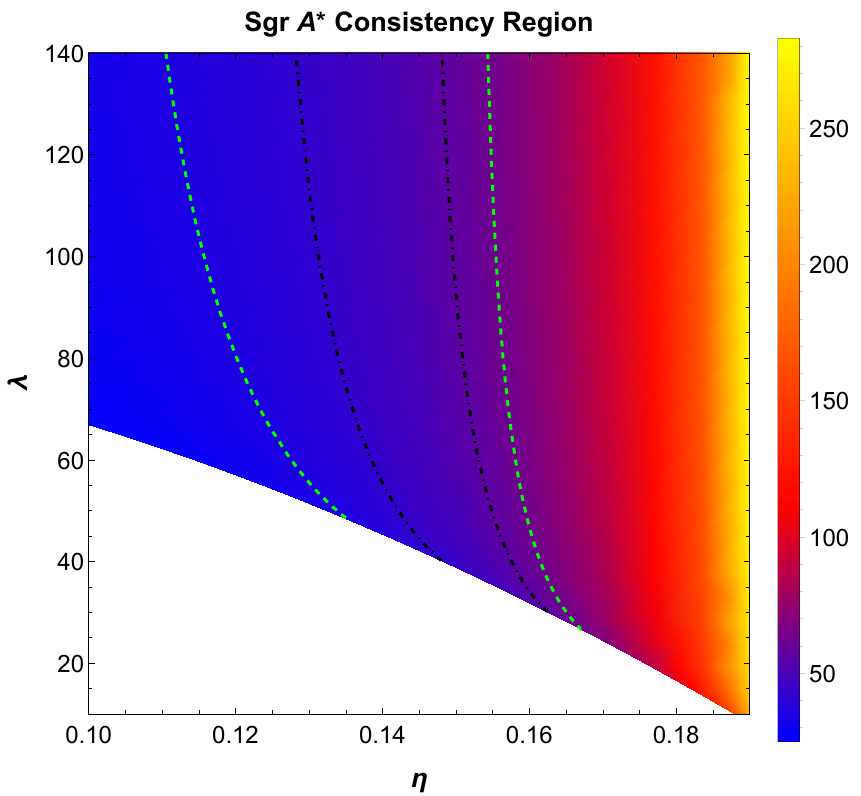}}\label{fig SgrA}}
   \caption{Consistency regions for the black hole angular shadow diameter in $\mu as$, in the $(\eta, \lambda)$ parameter space, evaluated against EHT observational data for M 87$^*$ (left) and Sgr A$^*$   (right). The color gradient indicates the angular shadow diameter in $\mu\text{as}$,
   while the black dashed lines and green dashed lines represent $1\sigma$ and $2\sigma$ confidence intervals derived from the observed results.  }
   \label{fig EHT}
\end{figure}

To compare the predictions of our model with the Event Horizon Telescope (EHT) observations, the dimensionless shadow radius $\mathcal{R}_s(\eta,\lambda)$ obtained from Eq.~(\ref{eqn sha}) is converted into the observable angular shadow diameter. Restoring physical units, the physical shadow radius is given by
\begin{align}
R_{\rm sh}^{\rm phys}=\mathcal{R}_s\frac{GM}{c^2},
\end{align}
where $M$ is the black hole mass. 
For a source at a distance $D$, the corresponding angular shadow diameter is
\begin{align}
d_{\rm sh}(\eta,\lambda)
=
2\mathcal{R}_s(\eta,\lambda)\frac{GM}{c^2D},
\end{align}
where the resulting angle is converted from radians to microarcseconds using the standard conversion factor. For convenience, we define the angular conversion factor
\begin{align*}
\mathcal{C}
=
\frac{GM}{c^2D}\times206265\times10^{6},
\end{align*}
so that
\begin{align}
d_{\rm sh}(\eta,\lambda)
=
2\mathcal{R}_s(\eta,\lambda)\mathcal{C}.
\end{align}
 In the present analysis, we adopt $M=6.5\times10^9\,M_\odot$ and $D=16.8\,{\rm Mpc}$ for M87$^*$, and $M=4.0\times10^6\,M_\odot$ and $D=8.0\,{\rm kpc}$ for Sgr A$^*$ \cite{Akiyama2019,Akiyama2022}. Throughout this analysis, the black hole mass and distance are fixed to their EHT central values for each source. Consequently, the degeneracy discussed below refers only to the model parameters $(\eta,\lambda)$, while possible correlations with the observational uncertainties in the black hole mass and distance are not considered. 
The EHT collaboration reported angular shadow diameters of $42\pm3~\mu{\rm as}$ for M87$^*$ and $48.7\pm7~\mu{\rm as}$ for Sgr A$^*$ \cite{Akiyama2019,Akiyama2022}. To determine the observationally allowed parameter space, the theoretical angular shadow diameter is required to satisfy
\begin{align}
d_{\rm lower}
\le
d_{\rm sh}(\eta,\lambda)
\le
d_{\rm upper},
\end{align}
where $d_{\rm lower}$ and $d_{\rm upper}$ denote the lower and upper bounds of the corresponding $1\sigma$ and $2\sigma$ EHT confidence intervals. The resulting constraints are shown in Fig.~\ref{fig EHT}. The black  dot dashed and green dashed curves represent the $1\sigma$ and $2\sigma$ confidence boundaries, respectively, while the colour scale denotes the theoretical angular shadow diameter in $\mu{\rm as}$. Parameter combinations $(\eta,\lambda)$ lying between these confidence curves are observationally allowed with the current EHT measurements, whereas regions that do not admit a physically acceptable horizon or photon sphere are excluded.
Although the EHT shadow diameter provides an important observational consistency test, it alone cannot uniquely determine the two model parameters $(\eta,\lambda)$. Consequently, the regions shown in Fig.~\ref{fig EHT} should be interpreted as parameter combinations compatible with the current EHT observations rather than as a unique determination of the model parameters or definitive evidence of the model's physical viability.  Additional and complementary observational probes, including higher-order photon-ring properties from black hole imaging \cite{Gralla2019,Johnson2020}, QNM spectra from gravitational-wave observations \cite{Berti2009}, and black hole polarimetric measurements \cite{EHTPol2021,EHTPol2024}, would be required to further constrain the parameter space and reduce this degeneracy.While the shadow diameter constrains the allowed parameter space through current EHT observations, it is also instructive to assess whether the strong-lensing observables themselves are likely to be detectable with present and future observational facilities.

\subsection{Observational Relevance and Detectability}
To assess the observational significance of the predicted strong-lensing signatures, we compare the calculated observables with the capabilities of current and prospective very-long-baseline interferometry (VLBI) facilities.  For an M87$^{*}$-like black hole, the EHT achieves an angular resolution of approximately $20\,\mu\mathrm{as}$, with image feature measurements reaching a precision of a few microarcseconds \cite{Akiyama2019,Akiyama2022}.
As shown in Table~\ref{tab lensing}, increasing $\eta$ from $0$ to $0.1$ increases $\theta_{\infty}$ and $\theta_1^{E}$ by approximately $14\,\mu\mathrm{as}$, suggesting that sufficiently large values of $\eta$ could, in principle, produce observable deviations comparable to the precision of current EHT measurements. In contrast, varying $\lambda$ from $105$ to $130$ changes $\theta_{\infty}$ by only about $1.47\,\mu\mathrm{as}$ (Table~\ref{tab lensing2}), which is challenging for current observations but may become testable with future high-resolution VLBI observations combined with advances in image reconstruction and parameter estimation techniques \cite{ngEHT2023,EHI2019}.

Nevertheless, the angular separation between successive relativistic images remains below $0.4\,\mu\mathrm{as}$ throughout the parameter range considered, which is well below the resolving capability of present interferometric facilities. Consequently, the individual relativistic images predicted by the strong-field lensing analysis are not expected to be directly observable with current instruments. Future VLBI arrays, including the next-generation Event Horizon Telescope (ngEHT) and proposed space-based interferometric missions, are expected to significantly improve the angular resolution and imaging fidelity, thereby enhancing the prospects for constraining the parameters $\eta$ and $\lambda$ through black hole imaging \cite{ngEHT2023,EHI2019}. The substantial increase in the relative time delay $\Delta T_{2,1}^{s}$ with increasing $\eta$ also provides an additional theoretical observable that could serve as a useful discriminator in future high-precision strong-lensing observations.

The above analysis of the deflection angle, lensing observables and shadow profile highlights the role of null geodesics in determining the optical appearance of the black hole. To gain further insight into the motion of massive particles in the spacetime, we extend the geodesic analysis beyond photon motion by studying the timelike geodesic equation and the associated effective potential.

\section{ Timelike Geodesics}
 The description of geodesics reveals the spacetime's basic characteristics and the behaviour of particles under gravitational influences \cite{NHeidari2024, RWang2024}. In this section, the timelike geodesic in a spherically symmetric black hole spacetime with global monopole and a coupling constant will be investigated. We analyze how the parameters $\eta$ and $\lambda$ influence the dynamics of test particles. 
For the metric  \eqref{metric}, the Lagrangian is written as \begin{eqnarray}\label{eqn Lagrangian}
2 \mathscr{L}= f(r) \dot{t}^2 - f(r)^{-1} \dot{r}^2- r^2 \dot{\theta}^2 - r^2 \sin^2\theta \dot{\phi}^2,
\end{eqnarray} where  the overdot denotes a derivative taken with respect to the proper time. Due to the spacetime's static nature and spherical symmetry, the geodesic motion will be considered in the equatorial plane, where $\theta={\pi}/{2}$ and $\dot{\theta}=0$. From the Lagrangian, the generalized momenta of the particle are derived as \begin{align}\label{eqn momenta}
p_t=f(r) \dot{t}=\mathcal{E}, \hspace{1cm} p_r=f(r)^{-1} \dot{r}, \hspace{1cm} p_\phi=-r^2 \dot{\phi}=-\mathcal{L},
\end{align}
where   $\mathcal{E}$  and $\mathcal{L}$ denote the conserved energy and angular momentum of the particle, respectively. Hence

\begin{eqnarray}
\dot{t}=\dfrac{\mathcal{E}}{f(r)},\hspace{0.5cm}
\dot{\phi}=\dfrac{\mathcal{L}}{r^2}.
\end{eqnarray} 
For timelike geodesics, using the above equations, Eq. \eqref{eqn Lagrangian} can be written as \cite{PriyoEPJC} 
\begin{eqnarray}
\dfrac{\mathcal{E}^2}{f(r)}-\dfrac{\dot{r}^2}{f(r)}-\dfrac{\mathcal{L}^2}{r^2}=1.
\end{eqnarray}
We get a first-order differential equation for the time-like geodesics with the conserved quantities $\mathcal{E}$ and $\mathcal{L}$ as \cite{BHamil2025}
\begin{eqnarray}\label{effective}
\dot{r}^2+V_{eff}(r)=\mathcal{E}^2, 
\end{eqnarray}
where the effective potential corresponding to the radial motion is derived as 
\begin{align}\label{eqn eff1}
V_{eff}(r)=f(r)\bigg(1+\dfrac{\mathcal{L}^2}{r^2}\bigg) \hspace{0.3cm}
= &
1+\dfrac{8 \pi G}{\lambda r^2}+\dfrac{\mathcal{L}^2}{r^2}+\dfrac{8 \pi G \mathcal{L}^2}{\lambda r^4}-8 \pi G \eta^2\bigg(1+\dfrac{\mathcal{L}^2}{r^2}\bigg) \nonumber\\ &-\dfrac{M}{r}-\dfrac{M \mathcal{L}^2}{r^3}.
\end{align}
Equation \eqref{effective} describes a system analogous to the one dimensional form of the equation of motion of a classical particle with energy $\mathcal{E}^2$ and effective potential  $V_{eff}(r)$ \cite{Ahmad2025}.
We remark that the effective potential can be defined in different ways depending upon how the radial equation is written. In the lensing section, we use $\dot{r}^2 = V_{ eff}(r)$, while in the geodesic analysis we adopt $\dot{r}^2 = \mathcal{E}^2 - V_{eff}(r)$. These two expressions are related by a simple redefinition of the effective potential and therefore the physical interpretation of the photon motion is not affected.

From Eq. \eqref{eqn eff1}, we see that the effective potential depends on multiple parameters including the global monopole parameter $\eta$, the coupling constant $\lambda$, the mass $M$ of the black hole and the angular momentum $\mathcal{L}$. The gravitational field of the black hole is modified by these parameters, consequently resulting in modifications in the test particles' motion. The effective potential reflects the impact on the motion of particles by the spacetime curvature. It provides significant physical insight of particle behaviour around a black hole. From Eq. \eqref{eqn eff1}, we see that as $r \rightarrow \infty $  , $V_{eff}(r) \rightarrow (1-8 \pi G \eta^2)$. The balance of the effective potential and the energy $\mathcal{E}$ will determine the particle's motion. Hence, by analyzing the effective potential's profile, the nature of different possible motions of the particles moving around the black hole can be discussed. If the total energy satisfies $\mathcal{E}^2 \geq(1-8 \pi G \eta^2)$, the particles may escape to infinity and the orbit which holds the condition is classified as unbound. If the total energy obeys $\mathcal{E}^2 <(1-8 \pi G \eta^2)$, the particle will obey a bound orbit \cite{BHamil2025}.
\begin{figure}[h!]
\centering
  \subfloat[\centering ]{{\includegraphics[width=170pt,height=170pt]{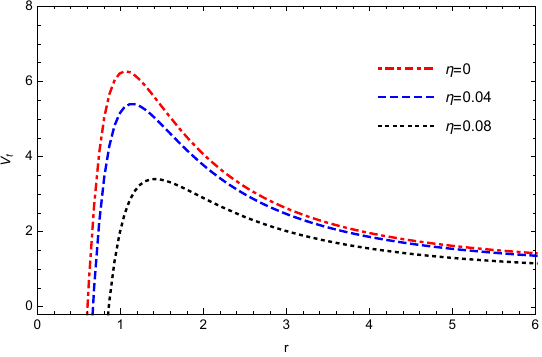}}\label{fig EPTL5eta}}
  \qquad
   \subfloat[\centering ]{{\includegraphics[width=170pt,height=170pt]{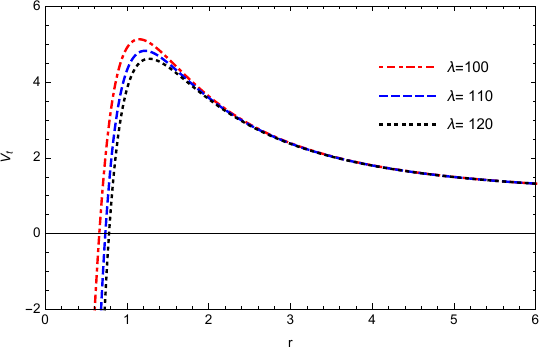}}\label{fig EPTL5lamb}}
\caption{Variation of the effective potential for different values of (a) $\eta$ with $\lambda=105$, $L=5$, $G=1$, $M=1$ and (b) $\lambda$ with $L=5$, $M=1$, $G=1$ and $\eta=0.05$.}
\label{fig EPtime1}
\end{figure}

\begin{figure}
\centering

\includegraphics[width=170pt,height=170pt]{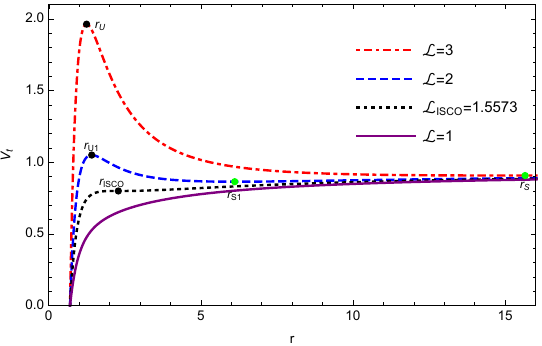}
\caption{Variation of the effective potential for varying $\mathcal{L}$ with  $M=1$, $\lambda=105$, $\eta=0.05$, $G=1$.}
\label{fig EPTLf}
 \end{figure}
 
In Figs. \ref{fig EPtime1} and \ref{fig EPTLf}, we illustrate the general behaviour of the effective potential $V_{eff}(r)$ as a function of the radial coordinate $r$. The unstable and stable circular orbits are determined by the maximum and minimum values of $V_{eff}(r)$. So, the unstable circular orbits can be inferred from  the effective potential graphs, where the maxima or peaks are associated with these orbits \cite{NabaJ2024}. From Fig. \ref{fig EPTL5eta}, we observe that increasing $\eta$ results in a reduction in the potential of the timelike geodesics. A lower effective potential barrier implies the particles require less energy to overcome the barrier, making escape trajectories more accessible. Moreover, we see that peaks shift to larger radial values with increasing $\eta$ for fixed $\lambda$ and $\mathcal{L}$ values which indicates that the radius of the unstable circular orbits is enlarged with increasing $\eta$. From Fig. \ref{fig EPTL5lamb}, we observe that the effective potential is reduced with increasing $\lambda$ for fixed $\eta$ and $\mathcal{L}$ values, suggesting a decline in the gravitational binding force experienced by a particle in orbit around the black hole. Moreover, the peak of the potential shifts outwards with increasing $\lambda$, indicating an outward shift of the circular orbits. \\
In Fig. \ref{fig EPTLf},  the effective potential is plotted with varying the angular momentum $\mathcal{L}$  and the maxima of the effective potential $(V_{eff}''(r)<0)$, corresponding to unstable equilibria are indicated by the black dots. The stable equilibria corresponding to the minimum potential for which $ (V_{eff}''(r)>0)$ are denoted by the green dots. We see that the maximum of the potential decreases with decreasing $\mathcal{L}$ and for $\mathcal{L}=1$ there is no maximum. This reveals that extremal points in the effective potential occur only when the angular momentum $\mathcal{L}$ exceeds $\mathcal{L_{ISCO}}$. Here, the unstable timelike geodesics will disappear for values of $\mathcal{L}$ less than $\mathcal{L}_{ISCO}$. Moreover, the curves showing an upward trend with increase in angular momentum values indicate that higher energy is needed to maintain the movement of the particles. We also notice that the radius of the unstable circular orbit decreases with increasing angular momentum $\mathcal{L}$, while the radius of the stable circular orbit increases. For $L=3$,  the unstable circular orbit occurs at $r_U=1.2483$, the stable circular orbit at $r_S=15.6586$ and for $L=2$, $r_{U1}=1.4179$ and $r_{S1}=6.1165$. The ISCO corresponding to $\mathcal{L}_{ISCO}=1.5573$ occurs at $r_{ISCO}=2.2909$ for the chosen values of the parameters.

\subsection{Circular orbits of timelike particles}

\begin{figure}[h!]
\centering
  \subfloat[\centering ]{{\includegraphics[width=170pt,height=170pt]{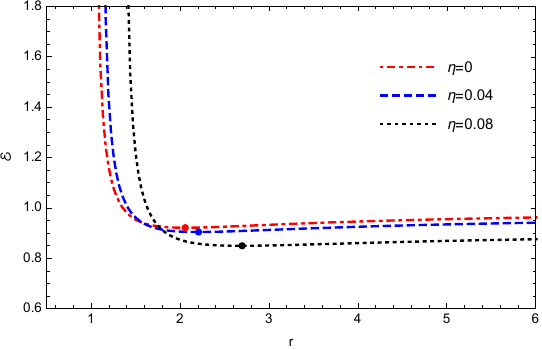}}\label{fig Eeta}}
  \qquad
   \subfloat[\centering ]{{\includegraphics[width=170pt,height=170pt]{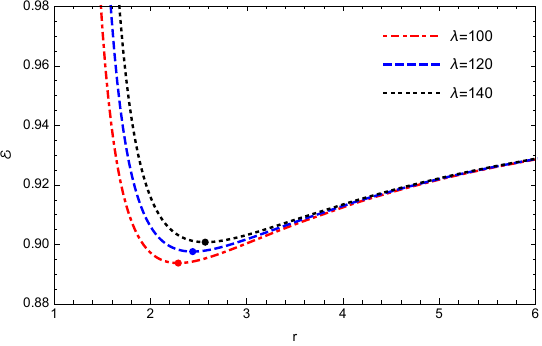}}\label{fig Elamb}}
   \caption{Radial profile of the energy $\mathcal{E}$ of timelike particles with respect to $\eta$ (left) and $\lambda$ (right). Here, (a) $M=1$, $\lambda=105$, $G=1$ and (b) $M=1$, $\eta=0.05$, $G=1$.}
   \label{fig EnergyE}
\end{figure}

\begin{figure}[h!]
\centering
  \subfloat[\centering ]{{\includegraphics[width=170pt,height=170pt]{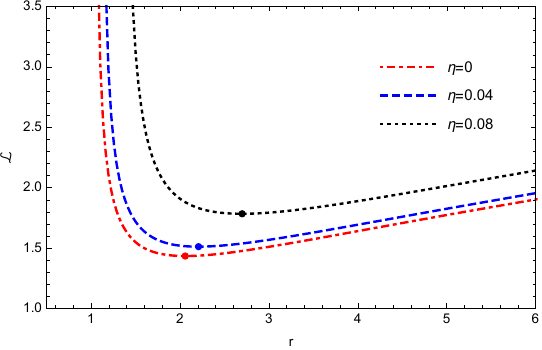}}\label{fig Leta}}
  \qquad
   \subfloat[\centering ]{{\includegraphics[width=170pt,height=170pt]{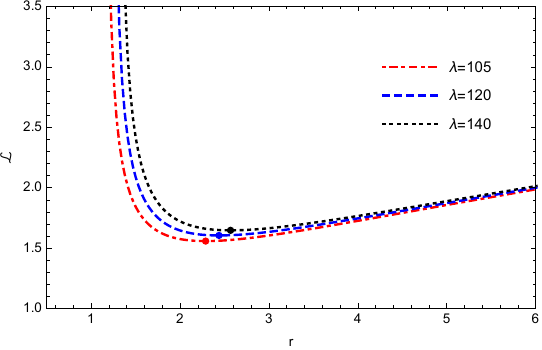}}\label{fig Llamb}}
   \caption{Radial profile of the angular momentum $\mathcal{L}$ of timelike particles with respect to $\eta$ (left) and $\lambda$ (right). Here, (a) $M=1$, $\lambda=105$, $G=1$ and (b) $M=1$, $\eta=0.05$, $G=1$.}
   \label{fig AngularL}
\end{figure}

For the analysis of the motion of timelike particles in the spacetime, the stable circular orbits and the ISCO are considered. A circular orbit will be maintained by a particle if two primary conditions are satisfied \cite{Tao-Tao2025, Izzet 2026 (5)}. The conditions are derived as
\begin{eqnarray}
&\dot{r}=0 \Rightarrow V_{eff}(r)=\mathcal{E}^2,\nonumber \\
&\ddot{r}=0 \Rightarrow V'_{eff}(r)=0,
\end{eqnarray}  
where $V'_{eff}(r)=\dfrac{\partial V_{eff}(r)}{\partial r}$.
Substituting Eq. \eqref{eqn eff1} into these conditions, we obtain the particles' energy and angular momentum corresponding to the circular orbits as
\begin{align}
\mathcal{L}^2&=\dfrac{16 \pi G r^2-M \lambda r^3}{16 \pi G \lambda \eta^2 r^2-2 \lambda r^2 -32 \pi G +3 M \lambda r},\\
\mathcal{E}^2&=\bigg(1-8 \pi G \eta^2+\dfrac{8 \pi G}{\lambda r^2}-\dfrac{M}{r}\bigg)\bigg(1+\dfrac{16 \pi G-M \lambda r }{16 \pi \lambda G \eta^2 r^2- 2 \lambda r^2 -32 \pi G + 3 M \lambda r}\bigg).
\end{align}
The second-order derivative of the effective potential decides the stability of these orbits. If 
$\dfrac{\partial^2 V_{eff}(r)}{\partial r^2}>0$, then the orbits is stable and if $\dfrac{\partial^2 V_{eff}(r)}{\partial r^2}<0$, it is unstable \cite{BHamil2025}. The radial dependence of the specific energy $\mathcal{E}$ is illustrated in Fig. \ref{fig EnergyE} for different values of the monopole parameter $\eta$ and parameter $\lambda$. From Fig. \ref{fig Eeta}, we see that as $\eta$ rises, the specific energy $\mathcal{E}$ decreases. Also, we find that the minimum value of energy is lowered by the presence of the global monopole. In contrast, we observe from Fig. \ref{fig Elamb} that the specific energy increases with the rise of the parameter $\lambda$. Furthermore, $\lambda$ lifts the minimum value of energy $\mathcal{E}$. Figure \ref{fig AngularL} displays the radial profile of the specific angular momentum $\mathcal{L}$, for varying values of $\eta$ and $\lambda$. The angular momentum decreases first and then increases with $r$. We notice that $\mathcal{L}$ increases with the rise of $\eta$ and $\lambda$. This indicates the particles orbit the black hole faster to maintain the circular orbit.  Also, the minimum value of $\mathcal{L}$ associated with the lowest point of the curve increases with increasing $\eta$ and $\lambda$, while shifting the corresponding inner circular orbit radius outward. This radius corresponds to $r_{ISCO}$, the radius of the ISCO. Therefore, from Figs. \ref{fig EnergyE} and \ref{fig AngularL}, we notice that the radius of the ISCO enlarges with the increase in the global monopole parameter $\eta$ and the constant parameter $\lambda$.  The coupling parameter $\lambda$ enters the effective potential through its contribution to the metric function, introducing additional radial dependent terms. Accordingly, the extrema of the effective potential are modified, shifting the positions of the stable and unstable circular orbits, consequently producing the observed changes in the ISCO radius.

The minimum radius at which stable circular orbits can exist is marked by the ISCO. Beyond the ISCO, the particles in circular motion become unstable, marking the boundary where the circular orbits shift from stable to unstable \cite{Shokhzod2025}. The ISCO is found by imposing the following conditions:
\begin{eqnarray}
V_{eff}(r)=\mathcal{E}^2, \hspace{0.6cm} \dfrac{\partial V_{eff}(r)}{\partial r}=0,\hspace{0.6cm} \dfrac{\partial^2 V_{eff}(r)}{\partial r^2}=0.
\end{eqnarray}
Substituting Eq. \eqref{eqn eff1} into the above condition, the ISCO constraint is established as 
\begin{eqnarray} \label{eqn isco constraint} 
r f''(r)f(r)+3 f(r)f'(r)-2rf'(r)^2\vert_{r=r_{ISCO}}=0.
\end{eqnarray}

\begin{table}[h]
 \centering
    \begin{tabular}{c| c c c c| c c c c}
    $\eta$   &  0 &   0.04 &  0.08 & 0.12 & 0.05 &  0.05 & 0.05 &0.05  \\  
\hline
$\lambda$ &   105 &   105 &  105 & 105 &   95 &  110 &  130 & 150\\
$r_{ISCO}$    &    2.06379 &   2.2066 &   2.69882  & 3.87892 &   2.14771  &  2.34722  & 2.5128 & 2.62239\\  

$\mathcal{L}_{ISCO}$  &   1.43368  &  1.51115 &   1.78324   & 2.45014 &   1.51316  &   1.5749   &   1.62727 & 1.66226\\
$\mathcal{E}_{ISCO}$  &   0.92061  &  0.90355 &   0.84877   & 0.74449 &   0.88964  &   0.89519  &   0.89933 & 0.90184\\
 \end{tabular}
\caption{Computed values $r_{ISCO}$, $\mathcal{L}_{ISCO}$ and $\mathcal{E}_{ISCO}$ for varying $\eta$ and $\lambda$ with fixed $M=1$ and $G=1$.}
    \label{tab shadow}
\end{table}

\begin{figure}[h!]
\centering
  \subfloat[\centering ]{{\includegraphics[width=170pt,height=170pt]{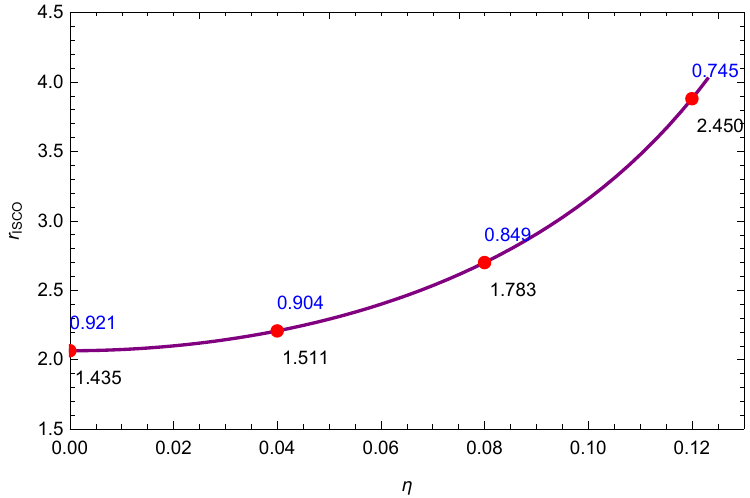}}\label{fig risco}}
  \qquad
   \subfloat[\centering ]{{\includegraphics[width=170pt,height=170pt]{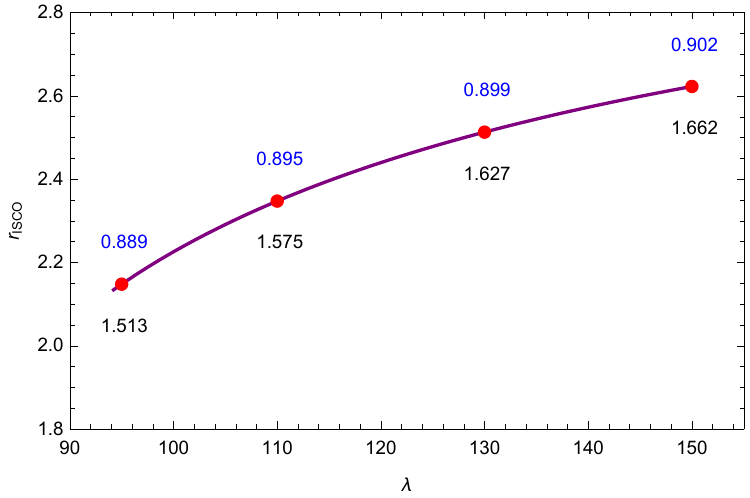}}\label{fig riscolamb}}
   \caption{Variation of the radius of the ISCO $(r_{ISCO})$ with respect to $\eta$ (left) and $\lambda$ (right). Here, (a) $M=1$, $\lambda=105$, $G=1$ and (b) $M=1$, $\eta=0.05$, $G=1$.}
   \label{fig RISCO}
\end{figure}

\begin{figure}[h!]
\centering
  \subfloat[\centering ]{{\includegraphics[width=170pt,height=170pt]{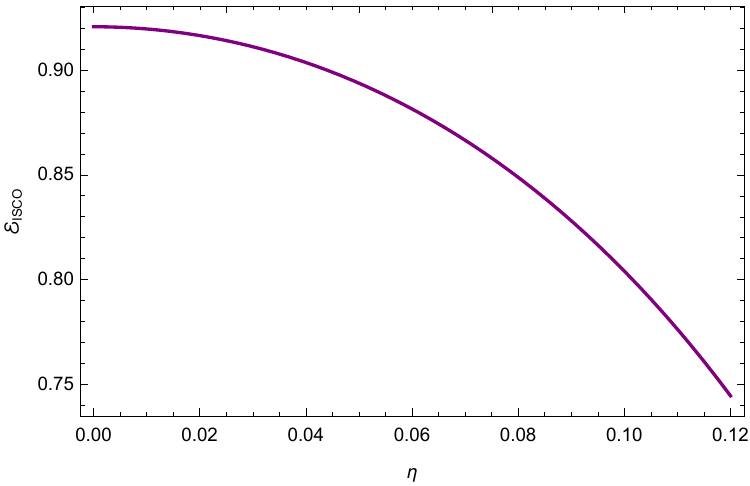}}\label{fig ESCOeta}}
  \qquad
   \subfloat[\centering ]{{\includegraphics[width=170pt,height=170pt]{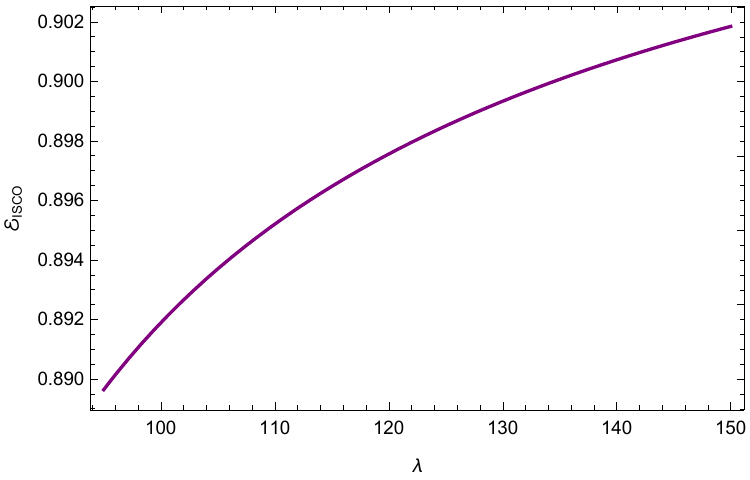}}\label{fig ESCOlamb}}
   \caption{Plot of the specific energy $\mathcal{E}_{ISCO}$ at ISCO with respect to $\eta$ (left) and $\lambda$ (right). Here, (a) $M=1$, $\lambda=105$, $G=1$ and (b) $M=1$, $\eta=0.05$, $G=1$.}
   \label{fig ESCO}
\end{figure}

\begin{figure}[h!]
\centering
  \subfloat[\centering ]{{\includegraphics[width=170pt,height=170pt]{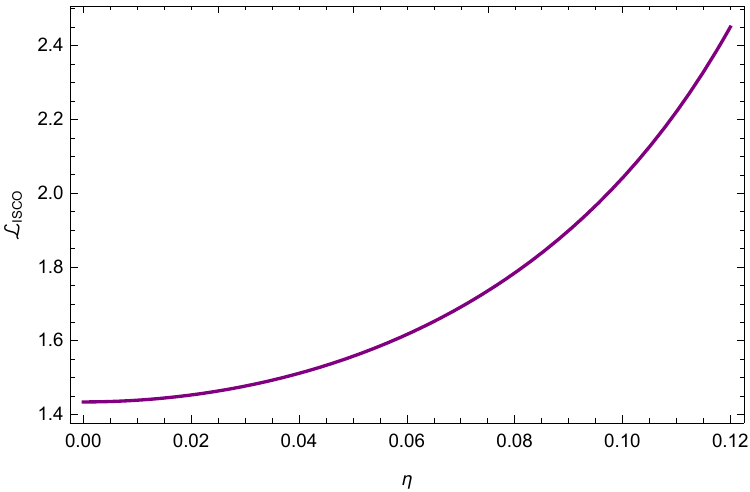}}\label{fig LSCOeta}}
  \qquad
   \subfloat[\centering ]{{\includegraphics[width=170pt,height=170pt]{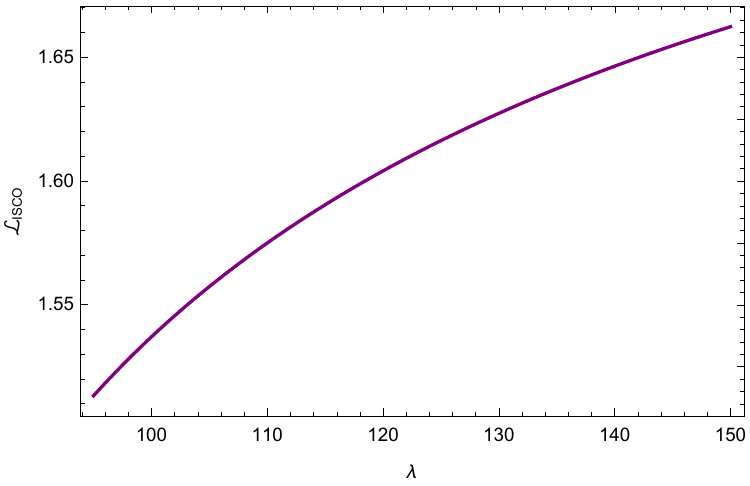}}\label{fig LSCOlamb}}
   \caption{Plot of the specific energy $\mathcal{L}_{ISCO}$ at ISCO with respect to $\eta$ (left) and $\lambda$ (right). Here, (a) $M=1$, $\lambda=105$, $G=1$ and (b) $M=1$, $\eta=0.05$, $G=1$.}
   \label{fig LSCO}
\end{figure}

Putting the expression of $f(r)$ in Eq. \eqref{eqn isco constraint}, we obtain:
\begin{eqnarray}\label{eqn risco}
M \lambda^2 r^3(1-8 \pi G \eta^2)-3M^2 \lambda^2 r^2+72 M \lambda \pi G r-512 \pi^2 G^2 \vert_{r=r_{ISCO}}=0.
\end{eqnarray}
The ISCO radius is obtained by solving Eq.\eqref{eqn risco} for the chosen values of the model parameters. In the limiting case $\eta\to0$ and $\lambda\to\infty$, the metric function reduces to $f(r)=1-\frac{M}{r}$, corresponding to the limiting solution of Rahaman et al. \cite{FRahaman2025}. Consequently, Eq.  \eqref{eqn isco constraint} consistently reproduces the ISCO radius $r_{\rm ISCO}=3M$. Since the mass parameter is normalized differently ($M=2M_{\rm Sch}$), this result is equivalent to the standard Schwarzschild value $r_{\rm ISCO}=6M_{\rm Sch}$.
In Fig. \ref{fig risco}, we plot the curves of $r_{ISCO}$ with respect to $\eta$ and $\lambda$, based on Eq. \eqref{eqn risco}. The numbers marked in blue and black represent the energy $\mathcal{E}_{ISCO}$ and angular momentum $\mathcal{L}_{ISCO}$ respectively of the corresponding orbit. We see that the ISCO radius, $r_{ISCO}$ increases monotonically with increasing $\eta$ and the rate of increase also rises with $\eta$. A larger ISCO radius means the region of stable circular motion moves outward, while a smaller ISCO indicates stable orbits closer to the black hole, showing that the unstable region contracts. From Fig. \ref{fig riscolamb}, we observe that $r_{ISCO}$ rises monotonically with $\lambda$, however, the rate of increase progressively diminishes with $\lambda$. It indicates that the effect of $\lambda$ on the orbital structure saturates at larger values of $\lambda$. Figures \ref{fig ESCO} and \ref{fig LSCO} illustrate the behaviour of energy $\mathcal{E}_{ISCO}$ and angular momentum $\mathcal{L}_{ISCO}$ at ISCO, varying with $\eta$ and $\lambda$. We see that  the specific energy at ISCO decreases due to the presence of $\eta$ while  it increases with $\lambda$, a behaviour consistent with Fig. \ref{fig EnergyE}. Moreover, we observe that the angular momentum corresponding to ISCO increases with $\eta$ and $\lambda$, though the effect is more pronounced at higher values of $\eta$. It means more angular momentum is  needed to allow a particle to remain in a stable orbit at ISCO. In Table \ref{tab shadow}, we show the ISCO radius $r_{ISCO}$ and the corresponding energy $\mathcal{E}_{ISCO}$ and angular momentum $\mathcal{L}_{ISCO}$, for different values of $\eta$ and $\lambda$. The tabulated values of $r_{ISCO}$, $\mathcal{E}_{ISCO}$  and $\mathcal{L}_{ISCO}$ show behaviour consistent with      the patterns observed in the graphs above. 

\subsection{Stability of the  timelike particles}

We will analyze the stability or instability of timelike geodesics in the spacetime using the Lyapunov exponents. The Lyapunov exponent evaluates how fast trajectories in the close vicinity of a spacetime either come together (converge) or move apart (diverge) over time \cite{NabaJ2024}. Here, we will find the Lyapunov exponents of massive particles in an unstable circular orbit on the black hole's equatorial plane. The Lyapunov exponent $\lambda_{\textrm{Ly}}$ is defined as \cite{BHamil2025} 
\begin{eqnarray}
\lambda_{\textrm{Ly}}= \sqrt{-\dfrac{1}{2 \dot{t}^2}\dfrac{\partial^2 V_{eff}(r)}{\partial r^2}}.\end{eqnarray}
By substituting the values of $\dot{t}$ and $V_{eff}(r)$ into the above expression, we have,
\begin{align}\label{lambdaL}
\lambda_{\textrm{Ly}}=&\dfrac{1}{\mathcal{E}}\bigg(1-8 \pi G \eta^2+\dfrac{8 \pi G}{\lambda r_c^2}-\dfrac{M}{r_c}\bigg)\bigg[ \dfrac{2 \mathcal{L}^2}{r_c^3}\bigg(\dfrac{M}{r_c^2}-\dfrac{16 \pi G}{\lambda r_c^3} \bigg)-\bigg(1+\dfrac{\mathcal{L}^2}{r_c^2}   \bigg) \bigg(\dfrac{24 \pi G}{\lambda r_c^4}-\dfrac{M}{r_c^3}    \bigg)   \nonumber\\ & 
- \dfrac{3 \mathcal{L}^2}{r_c^4} \bigg(1- 8 \pi G \eta^2+\dfrac{8 \pi G}{\lambda r_c^2}-\dfrac{M}{r_c}    \bigg)\bigg]^{\frac{1}{2}}.
\end{align}

\begin{figure}[h!]
\centering
  \subfloat[\centering ]{{\includegraphics[width=170pt,height=170pt]{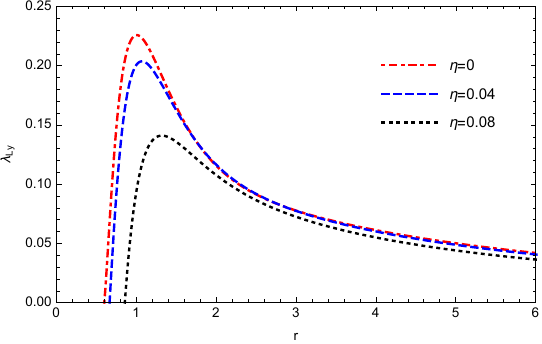}}\label{fig Lyaeta}}
  \qquad
   \subfloat[\centering ]{{\includegraphics[width=170pt,height=170pt]{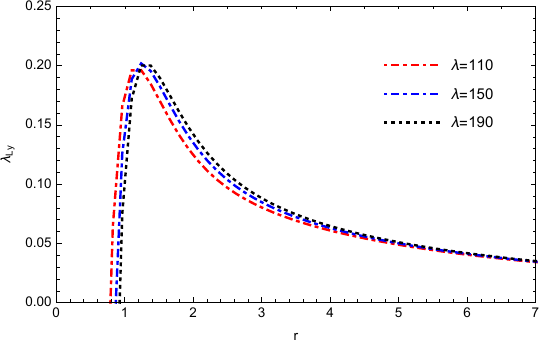}}\label{fig Lyalamb}}
   \caption{Illustration of the Lyapunov exponent $\lambda_L$ for varying  $\eta$ (left) and $\lambda$ (right). Here, (a) $M=1$, $\lambda=105$, $L=1$, $G=1$ and (b) $M=1$, $\eta=0.05$, $L=1$, $G=1$.}
   \label{fig Lyapunov}
\end{figure}

From Eq. \eqref{lambdaL}, we observe that several factors influence $\lambda_{\textrm{Ly}}$, including the global monopole parameter $\eta$ and the parameter $\lambda$.  The  stable or unstable nature of the circular orbits of timelike particles can be shown by the Lyapunov exponent $\lambda_{\textrm{Ly}}$ given above. The circular orbits are stable, unstable and marginally stable for complex nature (imaginary), real and zero values of $\lambda_{\textrm{Ly}}$ respectively \cite{Shobhit2022, RWang2024}. In Fig. \ref{fig Lyapunov}, we plot the radial variation of the Lyapunov exponent for varying values of $\eta$ and $\lambda$. From the figure, we see that $\lambda_{\textrm{Ly}}$ is positive and real for suitable values of the monopole parameter and $\lambda$, indicating unstable orbits in the range. Moreover, the instability decreases with the rise of the radius of the circular orbits and the orbits become marginally stable at higher values of the radius. It is also observed that increasing the global monopole parameter $\eta$ reduces the instability of the circular orbits and the unstable orbits existing region contracts. The decrease of $\lambda_{\textrm{Ly}}$ is also reflected in the effective potential plot, Fig. \ref{fig EPTL5eta}, where increasing $\eta$ decreases the height and sharpness of the potential barrier. Since the curvature of the potential at the peak governs the Lyapunov exponent, the reduced curvature or sharpness leads to decrease in $\lambda_{\textrm{Ly}}$, indicating a weaker instability. Increasing the parameter $\lambda$ makes the circular orbits more unstable.   

The analysis of circular orbits and their stability provide important information about particle dynamics in the given spacetime. To further probe the physical properties of the black hole, it is crucial to examine its response under external perturbations, which we consider in the following section.

\section{Electromagnetic perturbation using Teukolsky equation}
 The study of electromagnetic perturbations and QNMs provides important insight into the stability and response of the black hole, complementing the geometrical results obtained earlier.
In this section, we will discuss the electromagnetic perturbation of a static spherically symmetric black hole with a global monopole. In the Newman-Penrose formalism, the Maxwell equations  for the scalars $\Phi_i$ $(i=0,1,2)$ can be written as \cite{UKhanal, N. Ibohal, N. Ibohal2013}
\begin{align}\label{eqn m1}
&D\phi_1-\delta^* \phi_0= -\kappa \phi_2  +2 \rho \phi_1+(\pi-2\alpha)\phi_0,\nonumber \\
&D\phi_2-\delta^* \phi_1=(\rho -2\epsilon)\phi_2 + 2\pi \phi_1    -\lambda_{\textrm{NP}} \phi_0,\nonumber\\
&\delta \phi_1-\Delta \phi_0=-\sigma \phi_2+2 \tau \phi_1 +(\mu -2 \gamma)\phi_0    ,\nonumber\\
& \delta \phi_2- \Delta \phi_1=(\tau -2 \beta) \phi_2+ 2 \mu \phi_1    -\nu \phi_0. 
\end{align} Here $\alpha$,  $\tau$, $\rho$, $\beta$, $\lambda_{\textrm{NP}}$, $\kappa$,  $\epsilon$, $\mu$,  $\nu$, $\tau$, $\gamma$, $\sigma$ denote the spin coefficients and  $\Delta$, $D$, $\delta$ and $\delta^*$  represent the directional derivatives.
The null tetrad basis vectors corresponding to the metric \eqref{metric} are chosen as 
\begin{align}
& l^{\mu}=\left\lbrace \dfrac{1}{f(r)},1,0,0\right\rbrace,\nonumber \hspace{1.8cm}
n^{\mu}=\left\lbrace \dfrac{1}{2},\dfrac{f(r)}{2} ,0,0\right\rbrace,\nonumber\\ 
& m^{\mu}=\left\lbrace 0,0,\dfrac{1}{\sqrt{2}r},\dfrac{i }{\sqrt{2} r\sin \theta}\right\rbrace,\nonumber \hspace{0.4cm} 
 \overline{m}^{\mu}=\left\lbrace 0,0,\dfrac{1}{\sqrt{2}r},\dfrac{-i}{\sqrt{2}r\sin \theta}\right\rbrace.
\end{align} 
The directional derivatives  associated with the null basis vectors are defined as \cite{UKhanal} \begin{eqnarray}\label{eqn diderivatives}
D=l^\mu \partial_\mu= \mathscr{D}_0, \hspace{0.3cm} \Delta= n^\mu \partial_\mu = -\dfrac{\Delta_r}{2 r^2}~\mathscr{D}_0^{+},  \nonumber \\
\delta=m^\mu \partial_\mu= \dfrac{1}{\sqrt{2}r}~\mathscr{L}_0^{+}, \hspace{0.3cm}  \delta^*= \overline{m}^\mu \partial_\mu= \dfrac{1}{r\sqrt{2}}~\mathscr{L}_0,
\end{eqnarray}where
\begin{align*}
\mathscr{D}_n=\dfrac{\partial}{\partial r}+ \dfrac{i \omega r^2}{\Delta_r}+\dfrac{n}{\Delta_r}\dfrac{d \Delta_r}{dr},\hspace{0.4 cm}
\mathscr{D}_n^{+}=\dfrac{\partial}{\partial r}- \dfrac{i \omega r^2}{\Delta_r}+\dfrac{n}{\Delta_r}\dfrac{d \Delta_r}{dr},\\
\mathscr{L}_n^{+}=\dfrac{\partial}{\partial \theta}+ n \cot \theta-m \csc \theta,\hspace{0.4cm}
\mathscr{L}_n=\dfrac{\partial}{\partial \theta}+ n \cot \theta+m \csc \theta.
\end{align*}
The  non vanishing spin coefficients are given by 
\begin{eqnarray}\label{eqn spin}
& \mu =\dfrac{-f}{2 r} =\dfrac{-\Delta_r}{2 r^3} ,\hspace{0.3cm} \beta = \dfrac{\cot \theta}{2\sqrt{2r}}, \hspace{0.3cm}  \gamma =\dfrac{f'}{4}= \mu+ \dfrac{1}{4r^2}\dfrac{d \Delta_r}{dr} \nonumber \\
&  \rho = \dfrac{-1}{r},
 \hspace{0.3cm} \alpha= -\dfrac{\cot \theta}{2\sqrt{2r}}.
\end{eqnarray}
Using the Eqs. \eqref{eqn diderivatives} and \eqref{eqn spin} and  making the transformations $\phi_0=\Phi_0$, $ \phi_1=\dfrac{1}{\sqrt{2}r}\Phi_1$ and $\phi_2=\dfrac{1}{2 r^2}\Phi_2$, in \eqref{eqn m1}, we obtain \begin{align}
&\left( \mathscr{D}_0+\dfrac{1}{r}\right) \Phi_1= \mathscr{L}_1\Phi_0,\label{eqn a}\\
&\left( \mathscr{D}_0-\dfrac{1}{r}\right) \Phi_2= \mathscr{L}_0\Phi_1,\label{eqn b}\\
&\mathscr{L}_0^{+}\Phi_1=-\Delta_r\left( \mathscr{D}_1^{+}-\dfrac{1}{r}\right)\Phi_0,\label{eqn c}\\
&\mathscr{L}_1^{+}\Phi_2=-\Delta_r\left( \mathscr{D}_0^{+}+\dfrac{1}{r}\right)\Phi_1.\label{eqn d}
\end{align}
After eliminating $\Phi_1$ from Eqs. \eqref{eqn a} and \eqref{eqn c} and Eqs. \eqref{eqn b} and \eqref{eqn d}, we get differential equations involving $\Phi_0$ and $\Phi_2$ as
\begin{eqnarray} \label{eqn phi0}
\mathscr{L}_0^{+}\mathscr{L}_1 \Phi_0+ \Delta_r \mathscr{D}_1 \mathscr{D}_1^{+}\Phi_0-2 i \omega r \Phi_0=0,
\end{eqnarray}
\begin{eqnarray} \label{eqn phi2}
 \Delta_r \mathscr{D}_0^{+} \mathscr{D}_0\Phi_2+\mathscr{L}_0\mathscr{L}_0^{+} \Phi_2+ 2 i \omega r \Phi_2=0.
\end{eqnarray}
By taking $\Phi_0=  S_{+1}(\theta) R_{+1}(r)$ and $\Phi_2=  S_{-1}(\theta) R_{-1}(r)$, the radial parts of Eqs. \eqref{eqn phi0} and \eqref{eqn phi2} can be decoupled as 
\begin{align}\label{eqn radialR+}
\Delta_r \mathscr{D}_1 \mathscr{D}_1^{+} R_{+1}(r)-2 i \omega r R_{+1}(r)= \lambda_{\textrm{sep}} R_{+1}(r),
\end{align}
\begin{align}\label{eqn radialR_}
\Delta_r \mathscr{D}_0^{+} \mathscr{D}_0 R_{-1}(r)+2 i \omega r R_{-1}(r)= \lambda_{\textrm{sep}} R_{-1}(r),
\end{align}
where $\lambda_{\textrm{sep}}$ is the separation constant.  Using the condition $\Delta_r\mathscr{D}_{n+1}=\mathscr{D}_n \Delta_r$, Eq. \eqref{eqn radialR+}  takes the form
\begin{eqnarray} \label{eqn g}
\left(\Delta_r \mathscr{D}_0 \mathscr{D}_0^{+}-2 i \omega r\right)\Delta_r R_{+1}(r)= \lambda_{\textrm{sep}} \Delta_r R_{+1}(r).
\end{eqnarray}

\begin{figure}[h!]
\centering
  \subfloat[\centering ]{{\includegraphics[width=170pt,height=170pt]{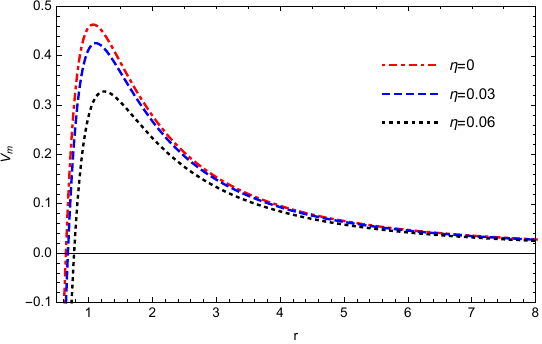}}\label{fig Epeta}}
  \qquad
   \subfloat[\centering ]{{\includegraphics[width=170pt,height=170pt]{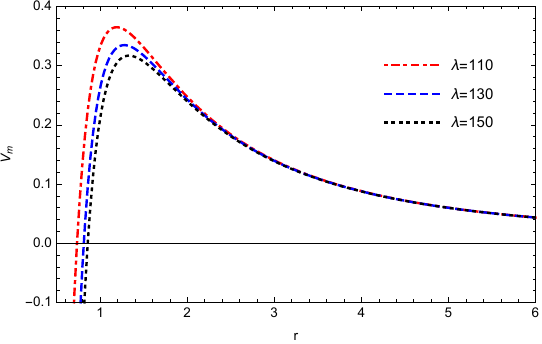}}\label{fig Eplamb}}
   \caption{Effective potential graph as a function of the parameters $\eta$ (left) and $\lambda$ (right).The physical parameters are chosen as (a) $M=1$, $\lambda=105$, $G=1$ and (b) $M=1$, $\eta=0.05$, $G=1$.}
   \label{fig Effective}
\end{figure}

 Eqs. \eqref{eqn g} and \eqref{eqn radialR_} can be written as
 \begin{align}\label{4.3}
  \bigg[\Delta_r^{-1}\dfrac{\partial}{\partial r}\Delta_r^{2}\dfrac{\partial}{\partial r}+\dfrac{1}{\Delta_r}\left(\omega^2 r^4+ i \omega r^2 \dfrac{d \Delta_r}{dr}\right)-4 i \omega r + \dfrac{d^2\Delta_r}{dr^2}-\lambda_{\textrm{sep}}\bigg]R_{+1}=0,
\end{align} and 
\begin{align}\label{4.4}
\left[\Delta_r \dfrac{\partial}{\partial r}\dfrac{\partial}{\partial r}+\dfrac{1}{\Delta_r}\left(\omega^2 r^4- i \omega r^2 \dfrac{d \Delta_r}{dr}\right)+4 i \omega r -\lambda_{\textrm{sep}}\right]R_{-1} =0.
\end{align}
Using the transformations $ R_{+1}=\dfrac{r Y_{+1}}{\Delta_r}$ and $R_{-1}=r Y_{-1}$, Eqs. \eqref{4.3} and \eqref{4.4}  take the following form
\begin{eqnarray}\label{4.5}
D^2 Y_{\pm 1} + P D_{\mp}Y_{\pm 1}- Q Y_{\pm 1}=0, 
\end{eqnarray} where $P$ and $Q$ are two functions, and $D_+$ and $D_-$ are two operators with 
\begin{align*}
D^2=D_+D_-= D_-D_+= \dfrac{d^2}{dr_*^2}+ \omega^2,\hspace{0.6 cm}
 D_{\pm}=\dfrac{d}{dr_*}\pm i \omega.
\end{align*} Here, $r_*$ is a generalized tortoise coordinate defined as $\dfrac{d}{dr_*}=\dfrac{\Delta_r}{r^2}\dfrac{d}{dr}$, where $\Delta_r=r^2 f$.
We also find that the new functions $P$ and $Q$ are established as \begin{align*}
& P=\dfrac{4 \Delta_r}{r^3}-\dfrac{1}{r^2}\dfrac{d \Delta_r}{dr}
 \hspace{0.3cm}= \dfrac{d}{dr_*}\ln  \mathscr{T}, \\
 & \mathscr{T}=\dfrac{r^4}{\Delta_r},  \hspace{0.3cm}
 Q=\dfrac{\Delta_r \lambda_{\textrm{sep}}}{r^4}.
\end{align*} 
Now further decomposing $Y_+$ as a linear combination of a function $Z$, we have   \cite{UKhanal} \begin{eqnarray} \label{4.6}
Y_+=f V_m Z +T D_+Z,\nonumber \\
D_-Y_+=G Z+h D_+Z,
\end{eqnarray}   Eq. \eqref{4.5} can be expressed as a one-dimensional Schr\"{o}dinger wave equation with respect to the coordinate $r_*$ as \begin{eqnarray}\label{eqn 4.8}
D^2 Z=V_m Z,
\end{eqnarray}  where $V_m$  denotes the potential barrier, provided the following system of equations are satisfied
\begin{eqnarray}\label{4.9}
\dfrac{d \mathscr{T}G}{dr_*}=\mathscr{T}( Q f V_m- h V_m), \\ \label{4.10}
\dfrac{\mathscr{T} h}{dr_*}=(QT+2 i \omega h -G)\mathscr{T},  \\\label{4.11}
h=\dfrac{dT}{dr_*}+fV_m,  \hspace{0.4cm} T=W+2i \omega f,\\\label{4.12}
G=\dfrac{d}{dr_*}(fV_m)+TV-2i\omega fV_m.
\end{eqnarray}
To prove that a solution $Z$ exists, which satisfies the Schr\"{o}dinger wave equation with the potential $V_m$, we have to solve the above set of equations.
We obtain the constant integral from the above four equations
\begin{eqnarray}\label{4.13}
\mathscr{T}(f V_m h-GT)=K,
\end{eqnarray} where $K$ is a constant.
Using Eq. \eqref{4.13} in Eq. \eqref{4.6}, we obtain
\begin{eqnarray}\label{4.15}
Z=\dfrac{\mathscr{T}}{K}(h-TD_-)Y_+,\\
\label{4.16}
D_+Z=\dfrac{\mathscr{T}}{K}(f V_m D_--D)Y_+.
\end{eqnarray}
To determine the explicit form of  $V_m$ appearing in Eq. \eqref{eqn 4.8}, we will take $T=2 i \omega$ which leads to $h=f V_m $ from Eq. \eqref{4.11}. Putting $\mathscr{T}h= \mathrm{constant}$, we derive the condition
\begin{eqnarray}\label{4.17}
\mathscr{T}Q=K'-\dfrac{A}{\mathscr{T}},
\end{eqnarray}
where $A=\dfrac{\mathscr{T}^2 h^2}{4\omega^2}$ and $K'=\dfrac{K}{4 \omega^2}+h\mathscr{T}$ are some constants. 
From Eq. \eqref{4.5} we have, $Q=\dfrac{\Delta_r \lambda_{\textrm{sep}}}{r^4}$, comparing with Eq. \eqref{4.17}, we identify $K'=\lambda_{\textrm{sep}}$ and $A=0$ which implies $\mathscr{T}h=0$. Therefore, from Eqs. \eqref{4.10} and \eqref{4.13}, we get $G\mathscr{T}=2i\omega \lambda_{\textrm{sep}}$ and $K=4 \omega^2\lambda_{\textrm{sep}}$.
Then, Eq. \eqref{4.9} gives 
\begin{eqnarray}
V_{m}=\dfrac{\Delta_r \lambda_{\textrm{sep}}}{r^4} = \dfrac{f(r) \ell(\ell+1)}{r^2},
\end{eqnarray}which is the required potential of electromagnetic perturbation for the given metric and $\lambda_{\textrm{sep}}$ denotes the separation constant given by $\lambda_{\textrm{sep}}=\ell(\ell+1)$.
In Fig. \ref{fig Effective}, we analyze the characteristic behaviour of  $V_m$ to explore how the parameters $\eta$ and $\lambda$ affect the black hole's spacetime structure. The parameters $\eta$ and $\lambda$ modify both the height and shape of the potential. We notice that increasing the parameters $\eta$ and $\lambda$ lowers the peak of the potential barrier, simultaneously exhibiting a narrower barrier shape. The pattern becomes more significant with the rise of the parameters $\eta$ and $\lambda$. A lower potential peak means the flow of waves to the surrounding medium is enhanced, which indicates that the greybody factors will rise with the parameters $\eta$ and $\lambda$. The observed modification of the potential barrier affects both the damping rates and oscillation frequencies of the    QNMs, which will be examined in the subsequent sections.  The potential barrier height has a direct correlation mainly with the oscillation frequency of the QNMs \cite{Jie2026}. A lower barrier usually corresponds to a lower oscillation frequency since the electromagnetic waves can escape more easily. The imaginary part of the QNMs is influenced by the potential barrier width. A wider barrier prolongs the damping period since it can trap the wave for an increased amount of time.

\section{Quasinormal Modes}

In this section, we will investigate the QNMs of the electromagnetic perturbation of the black hole considered. The QNM frequencies will be calculated numerically using WKB-Pad\'{e} approximation and the AIM method. The QNMs of electromagnetic perturbation arise as solutions of the wave equation \eqref{eqn 4.8} satisfying special boundary conditions:  purely ingoing at the event horizon and purely outgoing at spatial infinity.

\subsection{WKB-Pad\'{e} approximaton}
 Schutz and Will \cite{Schutz1985} first introduced the WKB method for calculating quasinormal frequencies, which was subsequently extended to third order by Iyer and Will \cite{Iyer1987a,S.IyerC.M.1987}. Later, Konoplya \cite{R.A.Konoplya2003} generalized the method to sixth order. To further improve the convergence and accuracy of the WKB approximation, Matyjasek and Opala \cite{J. Matyjasek2017,J. Matyjasek2019} proposed the Pad\'{e} resummation technique. In this work, we use the Pad\'{e}-averaged sixth-order WKB approximation to compute the quasinormal frequencies. The corresponding WKB formula is given by \cite{Yenshembam b2025}
\begin{eqnarray}
\dfrac{i(\omega^2-V_0)}{\sqrt{-2V''_0}}-\sum_{i=2}^6\Lambda_i=n+\dfrac{1}{2},
\end{eqnarray}
where $n$ denotes the overtone number and the subscript 0 denotes evaluation at $r_*(r_0)$, the location where the effective potential attains its peak.  $V''_0$ is the second derivative of the potential with respect to $r_*$  evaluated at $r_0$. The expressions of $\Lambda_i (i = 2, 3, 4, 5, 6)$ are given in \cite{S.IyerC.M.1987,R.A.Konoplya2003}. When the multipole number $l$ exceeds  $n$, the WKB method gives an accurate and reliable result. However,  when $n>l$, its accuracy decreases. To
obtain more reliable estimates of QNMs of higher order, Refs. \cite{J. Matyjasek2017,J. Matyjasek2019} proposed an extension of the WKB formalism by incorporating Padé approximation.


\subsection{AIM method}
 In addition to the above method, we also apply AIM to calculate the QNM frequencies \cite{Cho2010}. A comparison between the results obtained from both methods will be performed to validate our findings. The AIM has been extensively applied in diverse areas of physics, particularly in black hole perturbation theory and quantum mechanics, because of its computing efficiency and accuracy. Here, the QNM frequencies can be accurately calculated by expanding the perturbation equation's solution around a regular point and employing a recursive relation among successive derivatives. In this study,
the radial wave equation \eqref{eqn 4.8} will be solved using AIM to compute the QNM frequencies of the electromagnetic perturbation for varying $\eta$ and $\lambda$. By defining a new variable $u=\dfrac{1}{r}$,
 Eq. \eqref{eqn 4.8} becomes
\begin{eqnarray}\label{eqn AIM1}
p(u)^2 Z''(u)+p(u)p'(u)Z'(u)+(\omega^2-V(u))Z(u)=0,
\end{eqnarray} where 
\begin{eqnarray}
p(u)=u^2-M u^3-8 \pi G \eta^2 u^2+\dfrac{8 \pi G u^4}{\lambda}.
\end{eqnarray}Here, dash  represents derivative with respect to the radial coordinate $u$. The surface gravity is  given by \cite{I.Ablu2010, Y. Priyo2023, S. Christina 2021} \begin{eqnarray}
\kappa_i=\dfrac{1}{2}\dfrac{d f(r)}{dr}\vert_{r\rightarrow r_i}=-\dfrac{1}{2}\prod _{i\neq j}(u_i-u_j).
\end{eqnarray}
Using the new variable $u$, the tortoise coordinate $r_*$ can we rewritten as \begin{eqnarray}\label{eqn r*}
r_*=\int \dfrac{dr}{f(r)}=-\int \sum_{i=1}^4\dfrac{A_i}{u-u_i}du.
\end{eqnarray}
From Eq. \eqref{eqn r*}, we have
\begin{eqnarray}
1=\sum_{i=1}^4(A_i)\prod _{j\neq i}(u-u_j)) \hspace{0.6 cm}
\Rightarrow A_i=-\dfrac{1}{2 \kappa_i}.
\end{eqnarray}
Thus $r_*$ can be written as \begin{eqnarray}
r_*= \ln\bigg[\prod_{i=1}^4(u-u_i)^{\frac{1}{2 \kappa_i}}\bigg].
\end{eqnarray}
We define the wave function as \begin{eqnarray}\label{eqn Z(u)}
Z(u)= e^{i \omega r_*}\chi(u).
\end{eqnarray}
To scale out the divergent behaviour at the event horizon $r_1$, we choose  $\chi(u)=(u-u_1)^{-\frac{i\omega}{\kappa_1}}\xi(u)$. Here $\kappa_1 $ denotes the surface gravity at  $r_1$ and $u_1=\dfrac{1}{r_1}$. Inserting Eq. \eqref{eqn Z(u)}  along with the expression for $\chi(u)$ into Eq. \eqref{eqn AIM1}, we obtain the final equation as 
\begin{eqnarray}\label{eqn AIM chi''}
\xi''(u)=\lambda_0(u) ~\xi'(u)+s_0(u)~\xi(u),
\end{eqnarray}
where \begin{align}
\lambda_0(u)&=\dfrac{2 i \omega}{\kappa_1(u-u_1)}+\dfrac{2 i \omega-p'(u)}{p(u)},\nonumber \\
s_0(u)&=\dfrac{2 \omega^2}{\kappa_1 p(u)(u-u_1)}+\dfrac{\omega^2}{\kappa_1^2(u-u_1)^2}-\dfrac{i \omega}{\kappa_1(u-u_1)^2}+\dfrac{i \omega p'(u)}{p(u)\kappa_1 (u-u_1)}+\dfrac{\lambda}{p(u)}.
\end{align}

Now, differentiating Eq. \eqref{eqn AIM chi''} $n$ times yields \cite{Cho2010}
\begin{eqnarray}
\xi^{n+2}=s_n(u)~\xi(u)+\lambda_n(u)~\xi'(u),
\end{eqnarray}
where \begin{eqnarray}\label{eqn lambn sn}
&s_n&=s'_{n-1}+s_0\lambda_{n-1}, \nonumber \\
&\lambda_n&=\lambda'_{n-1}+\lambda_0 \lambda_{n-1}+s_{n-1}.
\end{eqnarray}
The improved AIM proposed in \cite{Cho2010} will be implemented. The functions $s_n$ and $\lambda_n$ are expanded around some point $\tilde{u}$ using the Taylor series as \cite{Cho2010} \begin{eqnarray}
&\lambda_n(u)&=\sum_{i=0}^{\infty} c_n^i(u-\tilde{u})^i, \nonumber \\
&s_n(u)&=\sum_{i=0}^{\infty} d_n^i(u-\tilde{u})^i.
\end{eqnarray}  Here $c_n^i$ and $d_n^i$ denote the $i^{th}$ Taylor expansion coefficients  of $\lambda_n$ and $s_n$ respectively. Using the above equation in Eq. \eqref{eqn lambn sn},  one obtains the following relations \cite{Yenshembam b2025}
\begin{eqnarray*}
&c_n^i&=(i+1)c_{n-1}^{i+1}+d_{n-1}^i+\sum_{k=0}^i c_0^k~c_{n-1}^{i-k},\\
&d_n^i&=(i+1)d_{n-1}^{i+1}+\sum_{k=0}^i d_0^k~d_{n-1}^{i-k}.
\end{eqnarray*}
The quantization condition is obtained as 
\begin{eqnarray}\label{eqn recurrence}
d_n^0~c_{n-1}^0-d_{n-1}^0~c_n^0=0.
\end{eqnarray}
The quantization condition given by Eq. \eqref{eqn recurrence} is solved numerically to obtain the quasinormal frequencies. In the improved AIM, the Taylor expansion is performed about the point $\tilde{u}=u_0$, where $u_0=1/r_0$ and $r_0$ denotes the location of the maximum of the effective potential. This choice provides the fastest convergence of the iterative procedure \cite{T. Barakat2005}. Throughout the present work, 50 AIM iterations are employed.
In Tables \ref{tab QNMeta} and \ref{tab QNMlamb}, we present the QNM frequencies for varying $\eta$ and $\lambda$ respectively. The real part of the QNMs corresponds to the actual oscillation frequency while the imaginary part is associated with the damping timescale and can be utilized to probe black hole stability. For a fixed overtone and multipole number, we observe that both the real and imaginary parts of the quasinormal frequencies decrease with increasing $\eta$ implying lower oscillation frequencies and slower damping. This behaviour suggests that larger values of the monopole parameter enhance the stability of the black hole.

  However, with increasing $\lambda$, the real part decreases monotonically but the absolute value of the imaginary part initially rises and then decreases for $\lambda=200$ except for $l=1$. This suggests that when $\lambda$ is small, the modes decay faster and it becomes slower for larger $\lambda$.  Further, we observe the following common behaviour in the two tables. For fixed values of $\eta$, $l$, $\lambda$, we find that the real part decreases while the magnitude of the imaginary part increases with larger $n$, indicating a lower oscillation frequency and a more rapid damping of the modes. However, for fixed $n$, both the real part and the magnitude of the imaginary part become larger with $l$. A larger $Re(\omega)$ corresponds to more oscillation of the electromagnetic wave, showing that it possesses more energy. Furthermore, we notice that as $l$ becomes larger, the absolute value of the imaginary part of the frequencies does not alter much, suggesting that the parameter $n$ mainly determines the attenuation rate of the quasinormal frequencies, instead of $l$.
    To quantitatively assess the consistency between the Pad\'{e}-averaged sixth-order WKB and AIM methods, we compare the RMS differences reported in Tables \ref{tab QNMeta} and \ref{tab QNMlamb}. For the representative fundamental mode $(l,n)=(2,0)$, the RMS differences are of the order of $10^{-6}$, while for the first overtone $(l,n)=(1,1)$, they remain below $10^{-3}$. These very small discrepancies demonstrate the excellent agreement between the two independent numerical approaches and confirm the numerical reliability of the reported quasinormal frequencies.

\begin{table*}[h]
\centering
\caption{Quasinormal frequencies $\omega$ obtained using the 6$^{\rm th}$-order WKB method with Pad\'e approximation and the AIM method for different values of  $\eta$ and fixed $\lambda=200$, $M=1$ and $G=1$.}
\begin{tabular}{c c c c c}
\midrule
$\eta$ & $(l,n)$ & WKB--Pad\'e &  AIM & $\Delta_{rms}$ \\ \midrule
\multirow{4}{*}{0.02}
& (1,0) & 0.547904 - 0.186059$i$	 & 0.547915 -0.186052$i$  & $9.22\times  10^{-6}$	\\
& (1,1) &  0.490935 - 0.585665$i$	 & 0.490921 -0.584526$i$	 &	$8.05\times  10^{-4}$  \\
& (2,0) & 	0.998873 - 0.19019$i$ & 0.998874 -0.190189$i$ &	$ 1 \times 10^{-6}$  \\
& (2,1) & 0.963547 - 0.579933$i$	 & 0.963550 -0.579880$i$  &	$3.75\times  10^{-5}$ \\ 
\midrule

\multirow{4}{*}{0.04}
& (1,0) & 0.522161 - 0.175015$i$ & 0.522170 -0.175011$i$ & $ 6.96\times  10^{-6}$ \\
& (1,1) & 0.468835 - 0.550354$i$  & 0.468887 -0.549398$i$ &  $ 6.77\times  10^{-4}$ \\
& (2,0) & 0.95076 - 0.178806$i$  & 0.950761 -0.178805$i$   & 1 $ \times  10^{-6}$ \\
& (2,1) & 0.917803 - 0.545028$i$  & 0.917801 -0.544982$i$   & $3.26 \times  10^{-5}$	\\ 
\midrule

\multirow{4}{*}{0.06}
& (1,0) & 0.480402 - 0.157333$i$ & 0.480409 -0.157330$i$&   $5.39 \times  10^{-6}$ \\
& (1,1) & 0.43301 - 0.493854$i$ & 0.433068 -0.493079$i$  & $5.5 \times  10^{-4}$ \\
& (2,0) & 0.872862 - 0.160592$i$ & 0.872863 -0.160591$i$  & $ 1\times  10^{-6}$	 \\
& (2,1) & 0.843669 - 0.489213$i$ & 0.843674 -0.489176$i$  & $ 2.64\times  10^{-5}$	 \\ 
\midrule

\multirow{4}{*}{0.08}
& (1,0) & 0.424319 - 0.134109$i$ & 0.424329 -0.134110$i$& $ 7.11\times  10^{-6}$   \\
& (1,1) & 0.384688 - 0.419836$i$ & 0.384759 -0.419295$i$ &  $3.86 \times  10^{-4}$ \\
& (2,0) & 0.768575 - 0.136706$i$ & 0.768575 -0.136705$i$  &	 $ 7.07\times  10^{-7}$ \\
& (2,1) & 0.744292 - 0.416067$i$ & 0.744293 -0.416044$i$  & $1.63 \times  10^{-5}$	 \\
\midrule
\end{tabular}  \label{tab QNMeta}
\end{table*}

\begin{table*}[h]
\centering
\caption{Quasinormal frequencies $\omega$ using the 6$^{\rm th}$-order WKB method with Pad\'e approximation and AIM methods for different $\lambda$ with fixed $\eta=0.05$, $M=1$ and $G=1$.}
\begin{tabular}{ccccc c}
\toprule
$\lambda$ & $(l,n)$ &  WKB--Pad\'e &  AIM & $\Delta_{rms}$ \\
\midrule

\multirow{4}{*}{130}
& (1,0) & 0.543401 - 0.166107$i$ & 0.543406 -0.166099$i$ & $ 6.67\times 10^{-6}$  \\
& (1,1) & 0.499627 - 0.517507$i$ & 0.499484 -0.516674$i$  & $ 5.98\times 10^{-4}$ \\
& (2,0) & 0.981192 - 0.169341$i$ & 0.981192 -0.169340$i$ & $7.07 \times 10^{-7}$ \\
& (2,1) & 0.954067 - 0.514535$i$ & 0.954056 -0.514498$i$ & $ 2.73\times 10^{-5}$  \\
\midrule
\multirow{4}{*}{150}
& (1,0) & 0.526385 - 0.167008$i$ & 0.526390 -0.167004$i$ & $4.53 \times 10^{-6}$  \\
& (1,1) & 0.479752 - 0.522467$i$ & 0.479715 -0.521456$i$ & $7.15 \times 10^{-4}$ \\
&(2,0) & 0.953452 - 0.17034$i$ & 0.953452 -0.170339$i$   & $7.07 \times 10^{-7}$ \\
& (2,1) & 0.924643 - 0.518262$i$ & 0.924633 -0.518217$i$ & $ 3.26\times 10^{-5}$  \\
\midrule
\multirow{4}{*}{180}
& (1,0) &  0.510411 - 0.167111$i$ & 0.510419 -0.167105$i$ & $ 7.07\times 10^{-6}$  \\
& (1,1) & 0.461057 - 0.524323$i$ & 0.461049 -0.523450$i$  & $ 6.17\times 10^{-4}$  \\
& (2,0) & 0.927253 - 0.17058$i$ & 0.927253 -0.170579$i$  & $7.07 \times 10^{-7}$ \\
& (2,1) & 0.896767 - 0.519568$i$ & 0.896767 -0.519527$i$ & $ 2.9\times 10^{-5}$  \\

\midrule
\multirow{4}{*}{200}
& (1,0) & 0.503192 - 0.166946$i$ & 0.503200 -0.166942$i$ & $ 6.32\times 10^{-6}$  \\
& (1,1) & 0.452582 - 0.524556$i$ & 0.452633 -0.523703$i$ & $ 6.04\times 10^{-4}$  \\
& (2,0) & 0.915353 - 0.170491$i$ & 0.915353 -0.170491$i$ & 0  \\
& (2,1) &  0.884116 - 0.519545$i$ & 0.884118 -0.519504$i$ & $2.9 \times 10^{-5}$  \\

\bottomrule
\end{tabular} \label{tab QNMlamb}
\end{table*}

Although QNM frequencies tell us the oscillation and damping properties, we also need to study how the perturbation evolves with time to fully understand the dynamical behaviour. This is addressed in the next section through a time-domain analysis.

\section{Evolution of electromagnetic perturbation around the black
hole}
\noindent
To further validate the QNMs spectrum obtained from the AIM and WKB--Pad\'e methods, we analyze the time domain evolution of electromagnetic perturbations. This approach provides a direct understanding of the system's dynamical response under external perturbations. To achieve this, a numerical integration scheme based on finite difference methods, originally developed in the context of black hole perturbations by Gundlach \textit{et al.} \cite{C. Gundlach1994} has been employed here.  We discretize the tortoise coordinate $r_*$ and the time $t$ on a uniform numerical grid, such that $\psi(i \Delta r_*,j \Delta t) = \psi_{i,j}$ and $V(r(r_*))=V(r_*,t)= V_{i,j}$.

The discrete form of the governing equation is expressed as
\begin{eqnarray*}
{\psi_{i+1,j} - 2\psi_{i,j} + \psi_{i-1,j}}{(\Delta r_*)^2} - \frac{\psi_{i,j+1} - 2\psi_{i,j} + \psi_{i,j-1}}{(\Delta t)^2} - V_i \psi_{i,j} = 0.
\end{eqnarray*} By rearranging the terms to solve for the future state $\psi_{i,j+1}$, we obtain the iterative evolution formula
\begin{align*}\psi_{i,j+1} = -\psi_{i,j-1} + \left( \frac{\Delta t}{\Delta r_*} \right)^2 \left( \psi_{i+1,j} + \psi_{i-1,j} \right) + \left[ 2 - 2 \left( \frac{\Delta t}{\Delta r_*} \right)^2 - V_i \Delta t^2 \right] \psi_{i,j}.\end{align*}
The initial perturbation is chosen as a Gaussian wave packet of width
$\sigma=2$ and center $k=10$, given by
\begin{eqnarray*}
\psi(r_*,0)=\exp\left[-\frac{(r_*-k)^2}{2\sigma^2}\right],
\end{eqnarray*}
with $\psi(r_*,t)=0$ for all $t<0$. The computational domain is chosen as
$r_*\in[-300,300]$ and discretized using $4000$ uniformly spaced grid points,
corresponding to a spatial step size
$\Delta r_*=\frac{600}{3999}\approx0.150$. The time step is taken as
$\Delta t=0.2\,\Delta r_*\approx0.030$, satisfying the Von Neumann stability
condition $\Delta t/\Delta r_*=0.2<1$. The waveform is recorded at the
midpoint of the computational domain ($r_*=0$). Outgoing-wave
(Sommerfeld-type) boundary conditions are imposed at both boundaries of the
computational domain to suppress spurious numerical reflections. The time
evolution is performed up to a final time of $t=200$. To verify numerical convergence, the simulations were repeated using a finer spatial grid consisting of $6000$ grid points while keeping all other numerical parameters fixed. The resulting time-domain profiles were found to be indistinguishable from those obtained with $4000$ grid points on the scale of the figures, confirming that the adopted grid resolution is sufficient for the present analysis.

\begin{figure}[h!]
\centering
  \subfloat[\centering ]{{\includegraphics[width=170pt,height=170pt]{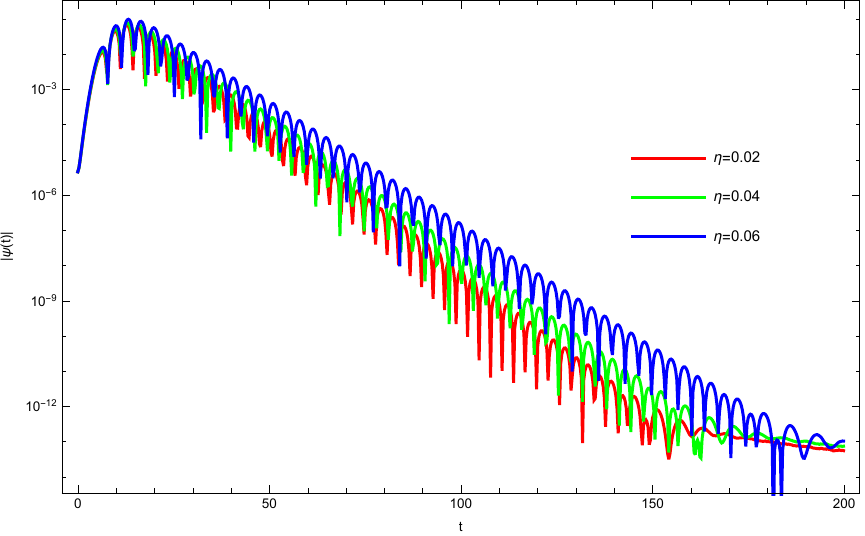}}\label{fig Ringdown_eta}}
  \qquad
   \subfloat[\centering ]{{\includegraphics[width=170pt,height=170pt]{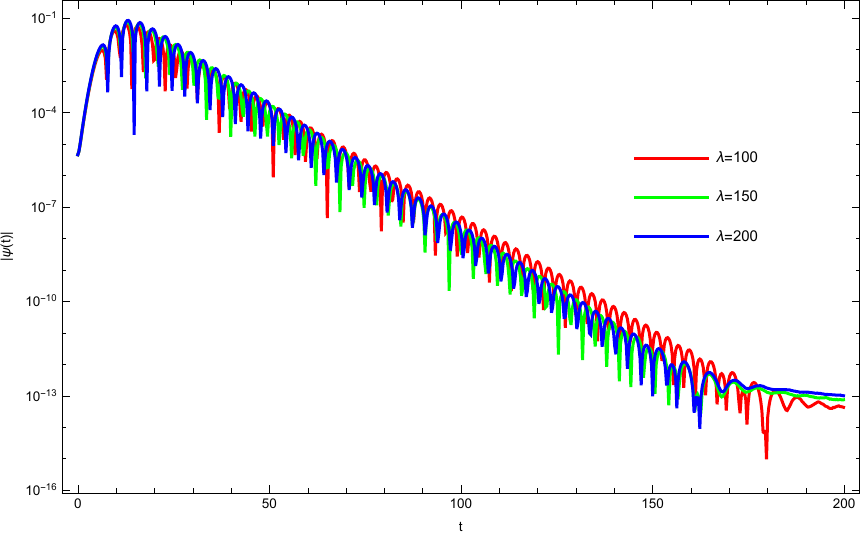}}\label{fig Ringdown_lambda}}
   \caption{Time-domain profiles of $|\psi(t)|$ (log scale) showing quasinormal ringing and late-time tails for different values of $\eta$ (left) and $\lambda$ (right).The physical parameters are chosen as (a) $M=1$, $\lambda=150$, $G=1$ and (b) $M=1$, $\eta=0.04$, $G=1$.}
   \label{fig Ringdown}
\end{figure}

In Fig. \ref{fig Ringdown}, we present the time-domain profiles of the perturbation field for different values of the monopole parameter $\eta$ and $\lambda$. The profiles are plotted in terms of $|\psi(t)|$ on a logarithmic scale as a function of time $t$, which effectively captures the exponentially decaying nature of the quasinormal ringing. The evolution clearly shows three separate phases: an initial burst, a quasinormal ringing stage and  finally a late-time tail. The intermediate ringdown phase is characterized by damped oscillations whose decay rate and oscillation frequency are governed by the imaginary and real parts of the quasinormal frequencies, respectively.
From Fig. \ref{fig Ringdown_eta}, it is evident that increasing the monopole parameter $\eta$ leads to a more gradual slope, indicating a decrease in the damping rate and allowing the perturbations to persist for a longer duration. Further from Fig. \ref{fig Ringdown_lambda}, it is observed that increasing $\lambda$ leads to a faster decay of the perturbation field, implying stronger damping. Consequently, the ringdown phase becomes shorter for larger values of $\lambda$. Since the time-domain evolution is presented only to illustrate the quasinormal ringing and late-time tails, no fitting procedure is applied to extract the quasinormal frequencies. The QNM frequencies reported in this work are obtained independently using the WKB--Pad\'{e} and AIM methods.

\section{Summary and Conclusion}

In this paper, we investigate a static and spherically symmetric spacetime with a global monopole, exploring the strong gravitational lensing effects, timelike geodesic structure, the dynamics of the test particles in the spacetime and its perturbation dynamics. In particular, we investigate the influence of the global monopole parameter $\eta$ on the deflection angle, the lensing observables and the shadow profile of the black hole. Furthermore, the impact of the monopole parameter on various physical quantities such as the metric function, effective potential, specific energy, specific angular momentum, ISCO is discussed. Additionally, we analyze the QNMs associated with electromagnetic perturbations, employing two different methodologies: the 6th order WKB method with Pad\'{e} approximation and the improved AIM method. Finally, we investigate the evolution of electromagnetic perturbations in the time domain profile.

It is noted that the Cauchy horizon and the event horizon exist when the condition given in Eq. \eqref{eqn realroots} is satisfied.  The two horizons exhibit dependence on the two characteristic parameters for monopole configuration, $\eta$ and $\lambda$, with higher $\eta$ and $\lambda$ increasing the event horizon radius. We notice that the event horizon radius is more sensitive to the change in $\eta$.
 The angle of deflection with global monopole is derived in Eq. \ref{defangle}.  We observe from Fig. \ref{def} that the deflection angle $\alpha(r_m)$ decreases with the rise of the impact parameter $b$ and it is found to diverge when $b \rightarrow b_c$. The behaviour for different values of $\eta$ and $\lambda$ is also shown. It is found that the divergent points increase for larger values of $\eta$ and $\lambda$ highlighting how the black hole parameters together with the global monopole, shape light bending. 
 
We also study the impact on lensing observables by the global monopole, both numerically and graphically.  We find that $\theta_\infty$ rises as we increase the two characteristic parameters for monopole configuration, $\eta$ and $\lambda$. However, the behavior of $s$ is not uniform across the two parameters. While $s$ is a monotonically increasing function of $\eta$, it diminishes monotonically with respect to $\lambda$. In the case of increasing $\eta$, $r_{\text{mag}}$ exhibits a non-monotonic behavior. It shows a slight initial increase followed by a subsequent decline whereas larger $\lambda$ values consistently yield a monotonic rise in $r_{\text{mag}}$. The two lensing coefficients, $\bar{a}$ and $\bar{b}$ respond quite differently to the parameters $\eta$ and $\lambda$. $\bar{a}$ is found to be a monotonically decreasing function of $\lambda$, whereas its behavior with respect to $\eta$ is non-monotonic: it first declines and subsequently rises. In contrast $\bar{b}$ shows a monotonic increase with both $\eta$ and $\lambda$. Both the angular radius of the outermost Einstein ring, $\theta^E_1$, and the time delay between the first and second relativistic images, $\Delta T^{s}_{2,1}$,  increase with rising $\eta$ and $\lambda$. The corresponding computed results are provided in Tables \ref{tab lensing} and \ref{tab lensing2}.

We also investigate the impact of the global monopole on the black hole shadow. The photon sphere radius and the corresponding shadow radius are derived, and we observe that increasing $\eta$ and $\lambda$ enlarges both the photon sphere and the shadow, with the effect being more pronounced for $\eta$. This indicates that the global monopole modifies the spacetime geometry in a manner that enlarges the region of unstable photon orbits, thereby increasing the apparent size of the black hole shadow. Furthermore, comparison of the theoretical angular shadow diameter with the EHT observations of M87$^*$ and Sgr A$^*$ identifies parameter combinations $(\eta,\lambda)$ that are consistent with the current observational measurements under the adopted black hole mass and distance. We also discuss the observational relevance and detectability of the predicted strong-lensing observables, showing that while the effects of $\eta$ may become accessible to future high-resolution VLBI observations, the relativistic images remain beyond the resolving capability of current instruments.

 For timelike geodesics,  we formulate the effective potential analytically for particles confined to the equatorial plane and investigate circular orbit characteristics including the ISCO, angular momentum and particle energy. The effective potential demonstrates  a clear dependence on $\eta$ and $\lambda$, with larger values of both the parameters reducing the height of the potential barriers. This implies that the gravitational influence is effectively weakened by the presence of global monopole, enabling more particle trajectories through the spacetime. We also find that the parameters $\eta$ and $\lambda$ significantly affect the stable circular orbits, orbital dynamics and the ISCO. For stable circular orbits, the dependence of the angular momentum $\mathcal{L}$ and the specific energy $\mathcal{E}$ on these parameters is clearly evident. We find that the specific angular momentum is enhanced for larger values of $\eta$ and $\lambda$. However, energy is reduced for larger $\eta$ while it is higher for larger $\lambda$,  demonstrating that the energetic properties of circular orbits are affected by the two parameters in distinct ways. We also observe that the parameters can modify the particle geodesics, thereby altering the radius of ISCO.  In our analysis, as illustrated by the graphs and table, we observe that the inclusion of the global monopole shifts the ISCO radius outward, suggesting  that the monopole induced modification shifts the stable circular orbits to larger radii. The increase in the ISCO radius is more pronounced for higher values of $\eta$ and the rate of increase becomes larger with $\eta$. This means a larger orbital radius is required for a stronger monopole field to maintain stability.  Moreover, it is revealed that the angular momentum and the energy at ISCO are also sensitive to change in the two parameters of the monopole configuration, as displayed via graphs.  These results show that the global monopole significantly modifies the orbital dynamics of massive particles, providing a possible means to distinguish these spacetimes from standard black hole solutions.\\
 In addition, we analyze the stability of timelike geodesic motion through the Lyapunov exponent to examine the stability of circular orbits for massive particles.  The Lyapunov exponent being positive for the considered parameter range indicates the instability of the circular orbit. It is revealed that the parameter $\eta$ reduces the instability of the orbits, while the parameter $\lambda$ enhances it slightly. Overall, the stabilizing influence of $\eta$ is much more pronounced than the destabilizing effect of $\lambda$.\\
  Moreover, we evaluate the QNMs of electromagnetic perturbation for varying $\eta$ and $\lambda$ using the 6th order WKB method with Pad\'{e} approximation and the AIM method. Our results reveal that both the global monopole parameter $\eta$ and the coupling constant $\lambda$ significantly influence the quasinormal frequencies within the range of parameters considered. In particular, for a fixed overtone and multipole number, we observe that the rise of the monopole parameter leads to weaker oscillation and slower damping of the quasinormal frequencies indicating that the effective stability of the spacetime is enhanced by higher values of the monopole parameter, making perturbations more persistent over time.  This behaviour can be attributed to the geometric modifications induced by the global monopole field. Furthermore, the variation of QNMs with $\lambda$ demonstrates the influence of the scalar field self coupling on the dynamical properties of the spacetime. The time-domain analysis of electromagnetic perturbations confirms the presence of the characteristic quasinormal ringing followed by late-time power-law tails. The observed dependence of the damping rate on the parameters $\eta$ and $\lambda$ is consistent with the  quasinormal frequencies determined by using the AIM method and the 6th order WKB method with Pad\'{e} approximation.

The results present a  systematic investigation of the interplay between the two characteristic parameters of monopole configuration $\eta$ and $\lambda$ and their implications on black hole physics.
Our study will contribute to a better understanding of how global monopole influences various observables of black holes, such as gravitational lensing, shadow profile and QNM frequencies, as well as the dynamics of test particles like the ISCO and stability. The present work excludes discussions on the different types of trajectories and orbits of the test particles in the spacetime.  We plan to address it in a subsequent study.  

\section*{Acknowledgements}

 The second author acknowledges the Council of Scientific and Industrial Research, New Delhi, India for giving financial support vide letter No. 211610101868. The authors also thank anonymous reviewers for valuable comments and suggestion to improve the paper.

\end{document}